# Polar Express: Rapid Functionalization of Single-Walled Carbon Nanotubes in High Dipole Moment Media

Dominik Just[1,*], Ryszard Siedlecki[1], Maciej Krzywiecki[2], Oussama Er-Riyahi[3], Yann Pouillon[3], Javier Junquera[4], Karolina Z. Milowska[3,5], Dawid Janas[1,*]

[1] Department of Chemistry, Silesian University of Technology, B. Krzywoustego 4, 44-100, Gliwice, Poland

[2] Institute of Physics – Centre for Science and Education, Silesian University of Technology, Konarskiego 22B, 44-100, Gliwice, Poland

[3] CIC nanoGUNE, Donostia-San Sebastián 20018, Spain

[4] Departamento de Ciencias de la Tierra y Física de la Materia Condensada, Universidad de Cantabria, Avenida de los Castros, s/n, E-39005 Santander, Spain

[5] Ikerbasque, Basque Foundation for Science, Bilbao 48013, Spain

* Corresponding author(s): dominik.just@polsl.pl, dawid.janas@polsl.pl

**Abstract:** Fluorescent semiconducting single-walled carbon nanotubes (SWCNTs) hold considerable promise for photonics. Furthermore, the optical characteristics of the material can be significantly improved by covalent modification, which generates new spectral features in the near-infrared region and enhances its photoluminescence quantum yield. However, despite the dynamic development of this research domain, the importance of the solvent environment in which the SWCNT functionalization is conducted remains relatively unexplored. In this work, the complex relationships between solvent, dispersant, and SWCNTs were untangled to unravel the underlying phenomena. Through a systematic investigation of SWCNT reactivity in a broad spectrum of solvents, supported by multi-scale modeling enabled by our new implementation of a hybrid functional within SIESTA, we discovered that both the solvent medium and the dispersant enabling SWCNT solubilization affect not only the kinetics but also the course of the covalent modification of SWCNTs. Polar solvents proved to induce significant structural reorganization of polymer molecules on the SWCNT surface and enhance charge redistribution at the polymer–SWCNT interface. Consequently, we achieved a high degree of control over the optical properties of SWCNTs, and the tailored SWCNTs enabled facile optical detection of cholesterol, a significant risk factor for cardiovascular diseases.

# 1. Introduction

Single-walled carbon nanotubes (SWCNTs) possess unique properties that captivate scientists, specifically in the context of photonics [1–6]. In particular, the sensitivity of SWCNTs' optical properties to the surrounding environment makes them promising for sensing a broad spectrum of analytes such as dopamine [7–9], serotonin [8], and cholesterol [10], among others [11–14]. However, the problem is that the photoluminescence quantum yield (PLQY) of these materials is unsatisfactorily low, hindering their utilization as sensors. Fortunately, light emission resulting from the radiative recombination of excitons (electron-hole pairs)[4] can be greatly improved by chemical modification of SWCNTs [5,15]. Consequently, the mobile excitons may be trapped at the as-generated defect sites, which function as potential energy wells [15]. This approach considerably enhances the photonic characteristics of SWCNTs. Firstly, the PLQYs of SWCNTs grafted with luminescent defects are often higher than those of pristine SWCNTs [4,5,15,16]. Secondly, the confinement of excitons creates new red-shifted PL bands, the emission of which is even better aligned with the transparency window of tissue. Therefore, optimizing the number and type of introduced luminescent defects to the SWCNT surface is essential.

These luminescent defects (also referred to as $sp^3$ defects or organic color centers (OCCs)) occur in various forms: (1) oxygen-based functional groups [10,17,18], (2) aliphatic substituents [19,20], and (3) aromatic moieties [19–23]. Several synthetic strategies have been designed to implant them, which give some degree of control over the optical emission characteristics of the covalently modified SWCNTs [4,5,20,24]. In addition to the aforementioned biomedical applications, which require near-infrared (NIR) emission, the PL at ca. 1,300 nm would be especially desirable. This way, the modified SWCNTs emitting in this range would also hold considerable value for telecommunication, as the 1,300 nm wavelength is among the most efficient bands for optical information technology [25]. Unfortunately, reaching this goal, especially using the most abundant (6,5) SWCNTs, is challenging. Commonly, the $E_{11}^{*-}$ (also called $E_{11}^{2*}$) peak positioned in this range emerges at an overly high SWCNT defect density, resulting in the deterioration of the optical properties of the material, such as PLQY [4,5]. Shiraki et al. obtained SWCNTs emitting in the vicinity of this range by using divalent bisdiazonium compounds [26]. This reactant was able to attach to two relatively close locations on the SWCNT, sufficiently disturbing the exciton recombination characteristics to obtain PL emission at ca. 1,250 nm. A similar concept was explored by Maeda et al., who used 1,4-diiodooctafluorobutane, a compound containing two reactive sites [27]. Due to two SWCNT attachment points and the presence of strongly electron-withdrawing fluorine substituents, the PL emission was registered at 1,320 nm. Furthermore, Settele and co-workers observed that 2-iodoaniline, when employed in the dark and attached to SWCNTs, also produced a new spectral feature at ca. 1,250 nm. This result likely originated from the dual attachment of 2-iodoaniline to SWCNTs. The importance of the configuration of the functional group added to the SWCNTs was confirmed by Wang et al., who revealed that not only the character of the aryl substituent, but also the chemical identity of the pairing group matters [28]. The use of various nucleophilic compounds that produce the pairing group can shift the PL emission band of (6,5) SWCNTs to nearly 1,300 nm. Lately, Yu and colleagues [29] as well as Maeda and his team [30] reported that the steric hindrance introduced by the substituent can also impact the electronic/optical properties of the SWCNTs to a sufficient extent, so that the newly generated PL peaks can be observed in the discussed spectral range. Last but not least, Espinoza et al. recently disclosed that the use of sodium hypochlorite under high pH conditions created a peak at 1,230 nm [31]. Despite the merits of these studies, the current understanding of the mechanism by which luminescent defects are introduced to SWCNT is still incomplete. This knowledge gap is partially caused by the lack of comprehension how the solvent environment and dispersant molecules affect the nature of SWCNTs as well as their propensity for chemical modification.

The impact of the solvent choice is rarely considered. Most commonly, SWCNTs are functionalized in water, which is an easy-to-handle medium, using reactive oxygen species [8,10,18,31,32] or diazonium chemistry [21,23,24,33]. Regarding organic solvents, which are less popular, two approaches dominate the field. Either SWCNTs are processed in toluene to implant an aryl group onto the surface [19,22,23], or they are subjected to reductive alkylation in anhydrous tetrahydrofuran [20,30]. Nonetheless, the state of the art lacks a detailed analysis of the impact of the solvent on the course of SWCNT functionalization. In this context, scientists have so far studied the influence of the solvent environment on the optical properties of pure [34] and already modified SWCNTs [34,35]. Recently, Heppe et al. investigated the solvent isotope effects between $H_2O$ and $D_2O$ when exposing SWCNTs to aryl-based diazonium salts [36]. The reaction of SWCNTs with aryl diazonium salts exhibited strong solvent dependency, suggesting that the surrounding medium significantly affects the process. The deposition of aryl groups onto the SWCNT surface generated a free radical on the SWCNT surface, which was paired with water-derived groups such as –OH or –OD. Furthermore, Berger and colleagues exploited the typically used toluene in conjunction with acetonitrile as a co-solvent, which, when employed with a potassium acetate (AcOK) as a base and an 18-crown-6 ether as a phase transfer agent, enabled solubilization of a diazonium salt to modify SWCNTs [23]. Unfortunately, the study of the impact of acetonitrile on the reaction course was beyond the scope of this work. Subsequently, the same group explored the possibility of adding up to 8.3% (each) of a mixture of tetrahydrofuran (THF) and dimethyl sulfoxide (DMSO) to modify SWCNTs with 2-haloanilines in the presence of potassium tert-butoxide (*t*-BuOK) base [23]. They concluded that THF speeds up the rate of the reaction by facilitating solubilization of the base in toluene. Additionally, it was deduced that DMSO enhances the basicity of *t*-BuOK, which increases the reactivity of the system. Hence, it is clear that the solvent environment may impact the course of SWCNT modification, but the origins of this phenomenon remain vague.

Concomitantly, one also needs to take into account the presence of the dispersing agent on the SWCNTs, which by themselves cannot be solubilized in water or typical organic solvents [32]. Yet, its possible role in the SWCNT functionalization has not been considered before to a sufficient extent. Typically, the consensus is that excess dispersant needs to be removed as it physically hampers SWCNT functionalization by restricting access to its surface [32], complicates assembly of SWCNT-based devices [24], or impedes charge transport within SWCNT networks [37]. Still, this step is not considered necessary for SWCNT functionalization when the dispersant concentration is low [22] or when it cannot encapsulate SWCNTs tightly [32]. To alleviate the aforementioned problems, Luo and co-workers performed direct functionalization of bulk unsorted SWCNT powders with diazonium salts generated *in situ* from the corresponding anilines in acidified water [38]. This method produced the same photoluminescence features in SWCNTs as if the material were first suspended in the liquid medium using a dispersant.

In this work, we disentangle the complex solvent-dispersant-SWCNT relationships, which are at the heart of SWCNT functionalization. To reach this goal, we capitalize on the benefits provided by phenylhydrazine (PhH) derivatives, which, as we recently reported, can be highly effective agents for SWCNT functionalization in both water and toluene [39]. Herein, we systematically employed a broad spectrum of solvents and process conditions, both of which were found to dramatically impact not only the kinetics but also the course of the covalent modification of SWCNT. As a result, both the reaction times were markedly shortened, and the spectral homogeneity of SWCNTs was very much improved. For example, in the case of the application of various xylenes (instead of toluene, typically used as the reaction medium), the reaction time needed to reach the same degree of SWCNT functionalization was decreased from 3 days to 60 min. Moreover, our multi-scale modeling (using a fresh implementation of hybrid functional within SIESTA) reveals that polar solvents such as acetonitrile and DMSO induce significant structural reorganization of PFO-BPy6,6' polymers and enhance charge redistribution at the polymer–SWCNT interface. These effects result in stronger electronic coupling, reduced band gaps, and partial hybridization of solvent and system states—particularly evident in DMSO—suggesting that solvent polarity actively modulates the electronic environment and facilitates functionalization of SWCNTs. These computational findings are consistent with our experimental observations, which show that polar media promote more efficient and selective defect formation. Consequently, the use of polar solvents enabled the exclusive generation of the most red-shifted $E_{11}^{*-}$ peak, which was registered at ca. 1,300 nm. Finally, the high utility of the covalently modified SWCNTs for photonics, prepared according to the described strategy, was demonstrated by illustrating how such materials can serve as high-performance optical nanosensors for cholesterol.

## 2. Results and Discussion

### 2.1. Functionalization of dispersant-free SWCNTs

Despite the recent surge of attention devoted to SWCNT functionalization to introduce luminescent defects into the material [4,5,20,24], the mechanism of the process remains elusive. To overcome this problem, it is essential to comprehend the interactions between all the entities present in the system (SWCNTs, dispersing agent used to suspend them, solvent environment, and reactive species introduced to be attached to SWCNTs). The influence of the solvent medium and dispersant on SWCNT functionalization appears to be underappreciated, which motivated us to analyze these aspects.

First, we investigated the impact of the polymer on the functionalization of SWCNTs by performing direct chemical modification of raw dispersant-free SWCNT powder [38] in various solvents (Figure 1a). Then, the functionalized SWCNTs were selectively solubilized using PFO-BPy6,6' to track the extent of chemical modification with (6,5) SWCNTs as the model material (Figure 1b). The spectrum of solvents may be divided into three groups: exclusively aromatic (benzene), alkyl-substituted aromatics, other aromatics, substituted aliphatics, and water. The application of alkyl-aromatics (toluene, xylenes, and p-cymene) resulted in the formation of distinct peaks in the optical absorption spectra (Figure 1b). Interestingly, in all these cases, the photoluminescence intensity peaked at ca. 1,120 nm, which corresponds to the relatively uncommon case of SWCNTs functionalized with alkyl groups ($E_{11\_(6,5)}^{*(AL)}$, Figure 1cd) [40]. It seems that in the absence of polymer, the generation of the benzyl radical from phenyl radicals produced by the decomposition of PhH was preferred. Benzyl radicals are extremely stable species, as they may be stabilized by multiple resonance structures, which increases the likelihood of SWCNT functionalization.

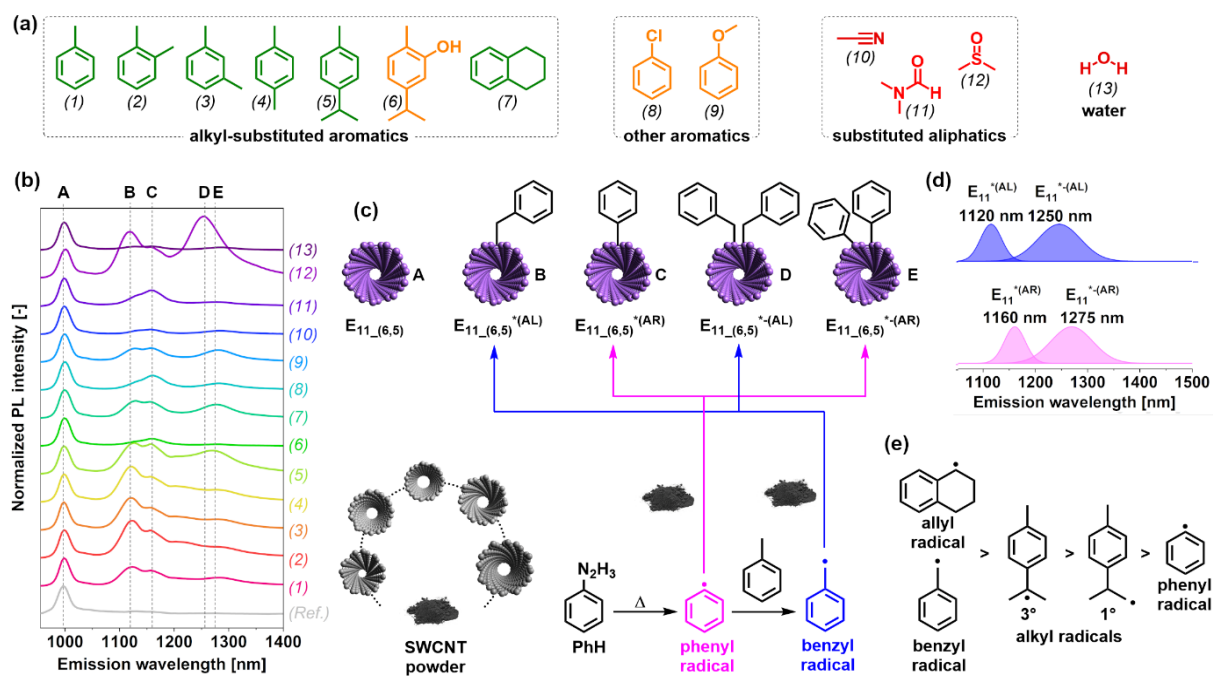

**Figure 1** (a) Various solvents used for chemical modification of SWCNTs with PhH divided into classes of compounds: (1) toluene, (2) *o*-xylene, (3) *m*-xylene, (4) *p*-xylene, (5) *p*-cymene, (6) carvacrol, (7) tetralin, (8) chlorobenzene, (9) anisole, (10) acetonitrile, (11) *N,N*-dimethylformamide (DMF), (12) dimethyl sulfoxide (DMSO), and (13) water. Non-polar, semi-polar and polar compounds are indicated in green, orange, and red, respectively, (b) PL spectra ($\lambda_{ex}$=579 nm) of raw SWCNT powder functionalized in the indicated solvents with PhH, (c) likely individual optical features (A-E) generated and observed in the PL spectra shown in panel (b) according to the indicated chemical reactions made possible by the treatment of polychiral SWCNTs with PhH in selected solvents (for the sake of clarity, radical transfer to solvent is only presented for toluene), which were subsequently extracted with PFO-BPy6,6' to isolate modified (6,5) SWCNTs. (d) Peaks observed for (6,5) SWCNTs grafted with alkyl (AL) and aryl (AR) moieties (low functionalization degree – left peak, high functionalization degree – right peak, (e) Most abundant radical species generated while functionalizing SWCNTs in a spectrum of solvents arranged in order of their stability.

Likewise, the exposure of PhH to SWCNTs in tetralin (aliphatic-aromatic compound) also produced an analogous peak, most probably due to the attachment of tetralin molecules to SWCNTs. Tetralin, upon exposure to phenyl radicals from PhH, may produce allyl radicals [41], which also exhibit remarkable stability (Figure 1e). In turn, these reactive centers may be the point of attachment to the SWCNT surface. In this case, the functionalization of SWCNTs with PhH was so effective that a relatively strong peak at ca. 1,250 nm was detected, which we ascribe to local modification of the SWCNT surface with multiple alkyl groups ($E_{11\_(6,5)}*\text{-}^{(AL)}$, Figure 1d). Moreover, both *p*-cymene and carvacrol contain an *i*-propyl group, which, after hydrogen abstraction, gives a tertiary radical, which is renowned for its stability as well, so it should give rise to improved chemical modification of SWCNTs in such media (Figure 1d). However, only in the former case were the $E_{11\_(6,5)}*^{(AL)}$ and $E_{11\_(6,5)}*\text{-}^{(AL)}$ features registered, while the latter medium hindered the modification of SWCNTs altogether. We presume that the inhibition of the SWCNT functionalization in carvacrol stems from the presence of a hydroxyl group in this compound, which is also found in many common radical scavengers [42]. This hypothesis is supported by the fact that the reaction in structurally similar anisole, for which the hydroxyl moiety (–OH) is protected by a methyl group (–OCH$_3$), proceeded to a similar extent as in the case of many other alkyl-substituted aromatics. The same reasoning explains why PhH was unable to modify SWCNTs in water.

Chlorobenzene was also evaluated as the solvent to verify our reasoning that benzyl radicals are made in the reaction of Ph• with the above-specified solvents, which may explain the formation of the peak at ca. 1,120 nm. Since this peak was not detected in the case of chlorobenzene, which lacks alkyl groups, this hypothesis was validated. From the quantitative standpoint, functionalization of SWCNTs in chlorobenzene was modest. A peak centered at ca. 1,160 nm was observed, corresponding to the attachment of aryl groups to the SWCNT surface ($E_{11\_(6,5)}*^{(AR)}$) as shown in Figure 1c [19,39,40,43,44]. Phenyl radicals generated by the decomposition of PhH are relatively unstable (Figure 1e). They cannot be stabilized by resonance because the radical orbital is aligned perpendicularly to the π-electron system. Since radical transfer from Ph• to solvent (chlorobenzene) would produce essentially the same reactive species, the extent of SWCNT modification (gauged by the intensity of the $E_{11}*$ peak to $E_{11}$ representing the unmodified SWCNTs) was not substantial.

Finally, we also tested three non-aromatic solvents often used in organic chemistry. In acetonitrile, the reaction did not progress, and the SWCNTs remained virtually untouched. Litwinienko *et al.* showed that the reaction rates of specific radical transformations are lower in acetonitrile and propionitrile [45] than when conducted in other exclusively hydrocarbon-based media, such as alkanes and benzene. Possibly, the reaction time or temperature was non-optimal for performing SWCNT functionalization in acetonitrile. On the other hand, the reaction conditions were sufficiently suitable for dimethylformamide, in which aryl groups were attached to the SWCNT surface, which was manifested by an $E_{11\_(6,5)}*^{(AR)}$ peak at ca. 1,160 nm. It should be noted that this feature is evident for many of the discussed solvents, meaning that the phenyl radicals initially generated from PhH were successfully attached to the SWCNT surface, as a considerable amount of them were neither subjected to recombination nor participated in radical transfer. Lastly, the processing of SWCNTs with PhH in DMSO yielded interesting results. Peaks at ca. 1,120 nm and 1,250 nm were detected, which align well with the previously discussed $E_{11\_(6,5)}*^{(AL)}$ and $E_{11\_(6,5)}*^{-(AL)}$ alkyl luminescent defects. However, the only alkyl groups in this compound system are in the solvent molecules, suggesting that the procedure may have resulted in the attachment of DMSO molecules to the SWCNT surface. Regardless of the chemical identity of the attached functional group, it is evident that SWCNT functionalization with PhH was most pronounced in DMSO. The intensity of the defect-induced $E_{11\_(6,5)}*^{(AL)}$ and $E_{11\_(6,5)}*-^{(AL)}$ peaks, were the highest among all solvents examined. Notably, the intensity of the latter exceeded the intensity of the native $E_{11}$ transition corresponding to fluorescence from unmodified parts of the SWCNTs. Characterization of the samples by Raman spectroscopy confirms that the highest $I_D/I_G$ ratio, a measure of the degree of SWCNT functionalization [46], was obtained for SWCNTs modified in DMSO (Figure S3). Considering all reaction media employed, DMSO is perhaps the strongest Lewis base. The Gutmann donor number (DN), which gauges Lewis acid or base strength, places DMSO (29.8) nearly on the same level as aniline (35.0) [47]. Concomitantly, we have recently demonstrated that PhH is most active toward SWCNT functionalization at a basic pH [39]. While the pH of DMSO cannot be determined, the basic conditions provided by the combination of DMSO [22] and PhH justify the effectiveness of SWCNT modification under these conditions.

### 2.2. Functionalization of polymer-wrapped SWCNTs

The modification of (6,5) SWCNTs with PhH progressed differently in the presence of PFO-BPy present on their surface (Figure 2a). Firstly, the recorded optical absorption spectra show that the peaks ($E_{11\_(6,5)}*^{(AL)}$) corresponding to the chemical attachment of the alkyl-bearing aromatic solvents positioned at ca. 1,120 nm disappeared (Figure 2b). Hence, contrary to previous reports [22,23,29,31,39,40,44,48], the polymer molecules residing at the SWCNTs not only hinder SWCNT functionalization, but these macromolecules also can alter the reaction course. As a result, the PL spectra were dominated by the $E_{11\_(6,5)}*^{(AR)}$ peaks located at ca. 1,160 nm, coming from the direct attachment of the phenyl groups to the SWCNTs from PhH degradation. Consequently, the polymer-coated SWNCTs favor functionalization with phenyl groups generated by PhH decay, and the effect of radical transfer to solvent is minimized.

Secondly, for the polar solvents such as anisole, acetonitrile, DMF, and DMSO, $E_{11\_(6,5)}*-$ peaks beyond 1,200 nm were apparent, indicating a substantial modification of the SWCNT surface, which suggests that these SWCNT solvents favor SWCNT functionalization. We infer that this effect is due to the fact that these liquid media have a different capacity for solubilizing PFO-BPy, which, especially for large molecular weight fractions, suffers from solubility issues [53]. Consequently, the polymer chains assume different conformation in polar solvents, which leads to pronounced functionalization of SWCNTs (Figure 2c).

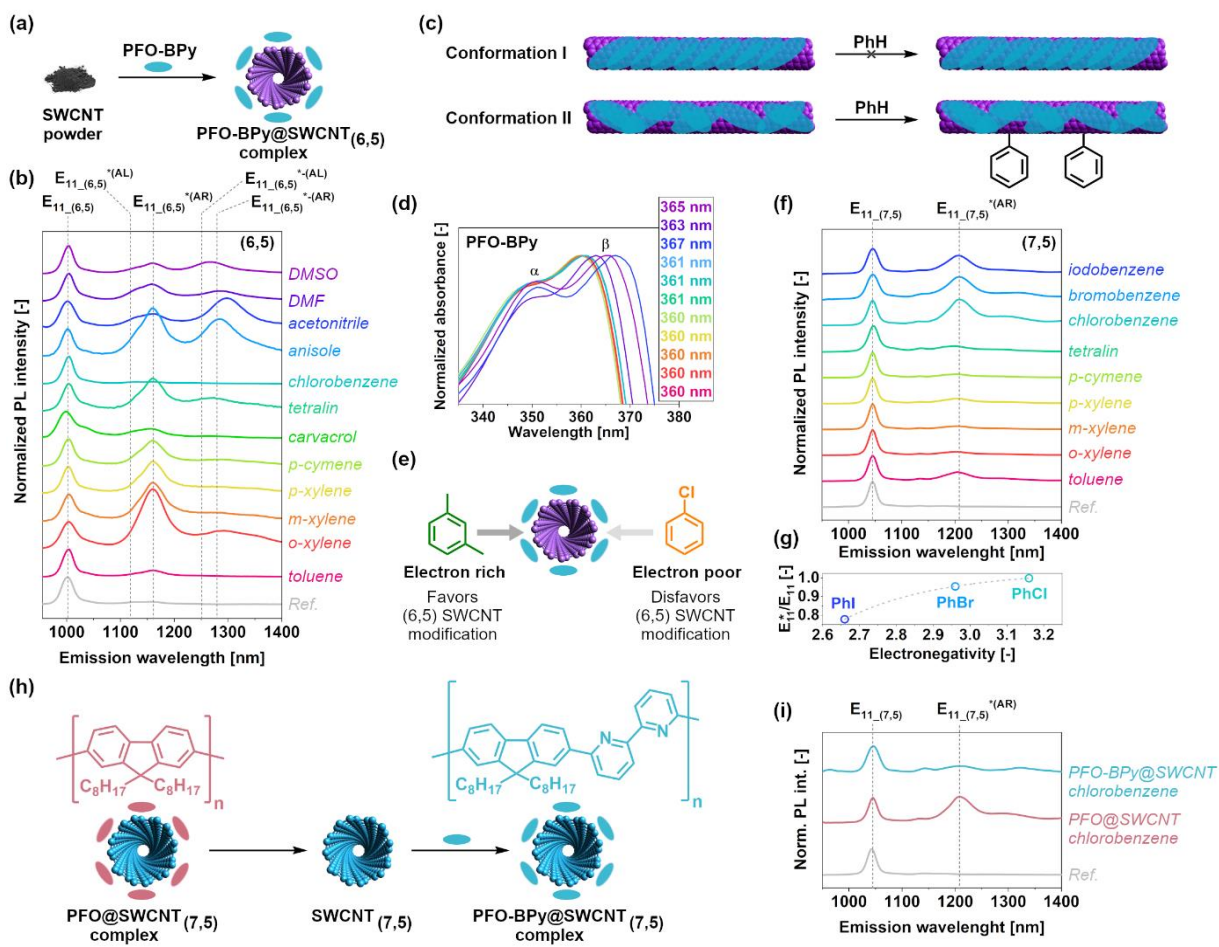

**Figure 2** (a) Selective extraction of (6,5) SWCNTs with PFO-BPy giving rise to the formation of the PFO-BPy@SWCNT complex, (b) PL spectra ($\lambda_{ex}$=579 nm) of (6,5) SWCNTs wrapped with PFO-BPy functionalized in the indicated solvents with PhH, (c) Schematic arrangement of polymer molecules on SWCNTs in 2 solvents leading to chemical modification of the material and the lack thereof, respectively, depending on the assumed conformation, (d) Optical absorption spectra of PFO-BPy in the examined solvents along with the positions of absorption maxima of the β phase, (e) Improved reactivity of (6,5) SWCNT encapsulated by PFO-BPy in electron rich solvents, (f) PL spectra ($\lambda_{ex}$=642 nm) of (7,5) SWCNTs wrapped with PFO functionalized in the indicated solvents with PhH, (g) the relationship between $E_{11}^*/E_{11}$ ratio of the covalently modified (7,5) SWCNTs and the electronegativity of the halogen atom attached to the benzene ring, (h) replacement of the polymer wrapping on (7,5) SWCNTs from PFO to PFO-BPy, and (i) PL spectra ($\lambda_{ex}$=642 nm) of (7,5) SWCNTs wrapped with PFO and PFO-BPy functionalized in chlorobenzene with PhH.

This reasoning is supported by the results of optical characterization of the examined polymers in various solvents (Figure 2d). For the polar solvents analyzed, the PFO-BPy peak exhibits improved spectral separation, wherein the optical features related to α- and β-phase [49–52] of the polymer are better resolved. The relative change of intensity of these peaks and dissimilarities in optical absorption wavelengths support the possibility of polymer rearrangement. This phenomenon will be analyzed in greater detail with modeling in subsequent parts of this work. Thirdly, comparison of how the alkyl-containing aromatic solvents affected the extent of SWCNT modification shows that the electron-rich xylenes and analogous alkyl derivatives of benzene performed better than toluene or chlorobenzene, which have a lower electron density. Thus, this observation, combined with the fact that the electron-rich polar solvents described above performed better, suggests that compounds of stronger Lewis base character are better suited for SWCNT modification. This conclusion is congruent with the results of the modification of raw polymer-free SWCNTs explained in the previous section (Figure 1). In both cases (with and without polymer on the SWCNT surface), relatively electron-deficient toluene and chlorobenzene were least potent for chemical modification of (6,5) SWCNTs, whereas the compounds containing a higher electron density (the polar solvents in particular) promoted SWCNT modification to the most significant extent.

Interestingly, the very same solvents (toluene and chlorobenzene) facilitated the chemical modification of the slightly larger (7,5) SWCNTs very well, which were selectively suspended with PFO (Figure 2f). (7,5) SWCNTs are notoriously hard to modify compared to the most commonly tailored (6,5) SWCNTs [4] as their increased diameter reduces the reactivity of the material due to lower curvature [53]. These unfavorable circumstances are further exacerbated by the fact that, to maintain the stability of the SWCNT dispersions, the thickness of the PFO coating surrounding (7,5) SWCNTs is much larger compared to the (6,5) PFO-BPy wrapping of (6,5) SWCNTs [54]. Still, distinct optical peaks at ca. 1,205 nm related to the modification of SWCNTs of this chirality with aryl groups ($E_{11\_(7,5)}*^{(AR)}$) were registered in the selected solvents. Based on this outcome, it can be concluded that the PFO@(7,5) SWCNT system exhibits a contrasting behavior to the PFO-BPy/(6,5) SWCNT one. Previously, we demonstrated that conjugated polymer/SWCNT complexes respond differently to solvents depending on their modality [55]. In brief, the molecules of conjugated polymers on the SWCNT surface exert strain on them, and the strain component is positive for mod = 1 SWCNTs, while it is negative for mod = 2 ones. Hence, we hypothesize that these contrasting characters of (6,5) (mod = 1) and (7,5) (mod = 2) SWCNTs, combined with the different behavior of the PFO-BPy and PFO, respectively, encapsulating them, may justify the observed effect. This justification is further supported by the results of Capaz et al. [56] and Kim et al. [57], who reported how SWCNTs of dissimilar modality behave differently.

To prove that the electron-poor solvents favor modification of (7,5) SWCNTs are not isolated incidents, (7,5) SWCNTs were exposed to PhH in a selection of halogen-containing aromatic solvents (chlorobenzene, bromobenzene, and iodobenzene). The more electron-deficient the system was, gauged by the electronegativity of the attached halogen atom, the more it favored (7,5) SWCNT modification (Figure 2g). The highest $E_{11}*/E_{11}$ ratio was recorded in chlorobenzene, for which the electronegativity of the chlorine atom was 3.16, compared to 2.96 and 2.66 for bromine and iodine, respectively [58–61].

Strikingly, the reactivity of (7,5) SWCNTs can also be controlled by the character of the conjugated polymer encapsulating them. The replacement of PFO, which was used for selective isolation of (7,5) SWCNTs with PFO-BPy (Figure 2h), strongly decreased the reactivity of (7,5) SWCNTs (Figure 2i). The $E_{11}*/E_{11}$ ratios of (7,5) SWCNTs coated with PFO and PFO-BPy, which were functionalized with PhH in chlorobenzene, reached 1.03 and 0.22, respectively. Pyridine is a Lewis base [62], so its presence in PFO-BPy contributes more electrons to (7,5) SWCNTs compared to PFO, which is a nitrogen-free polymer. The results of other experiments described above show that when (7,5) SWCNTs are in an electron-rich environment, their reactivity is moderated, which explains why the PFO-BPy@(7,5) SWCNT complex was much less prone to chemical modification than PFO@(7,5) SWCNT.

Apart from the chemical identity of the solvent or the polymer wrapping, radical reactions are susceptible to the presence of moisture [40]. Hence, we measured the water content of the examined solvents before and after drying them with 4 Å molecular sieves by means of Karl Fischer titration (Figure 3a). In most cases, the moisture level was relatively low and close to the detection limit of the used instrument. The highest water content was recorded for carvacrol and the polar solvents used. The former compound contains a hydroxyl group, which increases its hydrophilicity. The latter compounds, on the other hand, due to their chemical character, can readily interact with water. As a matter of fact, they are fully miscible with water [63], which should give rise to homogeneous distribution of water in these liquid media (no phase separation). Interestingly, the reactivity of PhH in the polar liquid media changed to a considerable extent depending on whether water was present or removed (Figure 3b). The most apparent difference was the increased preference for the formation of the $E_{11}*$ peak after drying, corresponding to a modest degree of implantation of luminescent defects, rather than $E_{11}*^-$, which is indicative of a substantial density of such defects. While molecular sieves are widely recognized for their ability to selectively adsorb moisture, their additional impact on the chemical environment has often been overlooked. They not only remove water but also impact the acidity of polar solvents [35,63,64], a phenomenon that critically affects our reaction pathways. Additionally, such sieves often also contain acidic or basic sites and can be used as acid/base catalysts [64]. The molecular sieves used in this work are composed of sodium aluminosilicate, which acts as a Lewis acid, altering the proton availability in certain solvents.

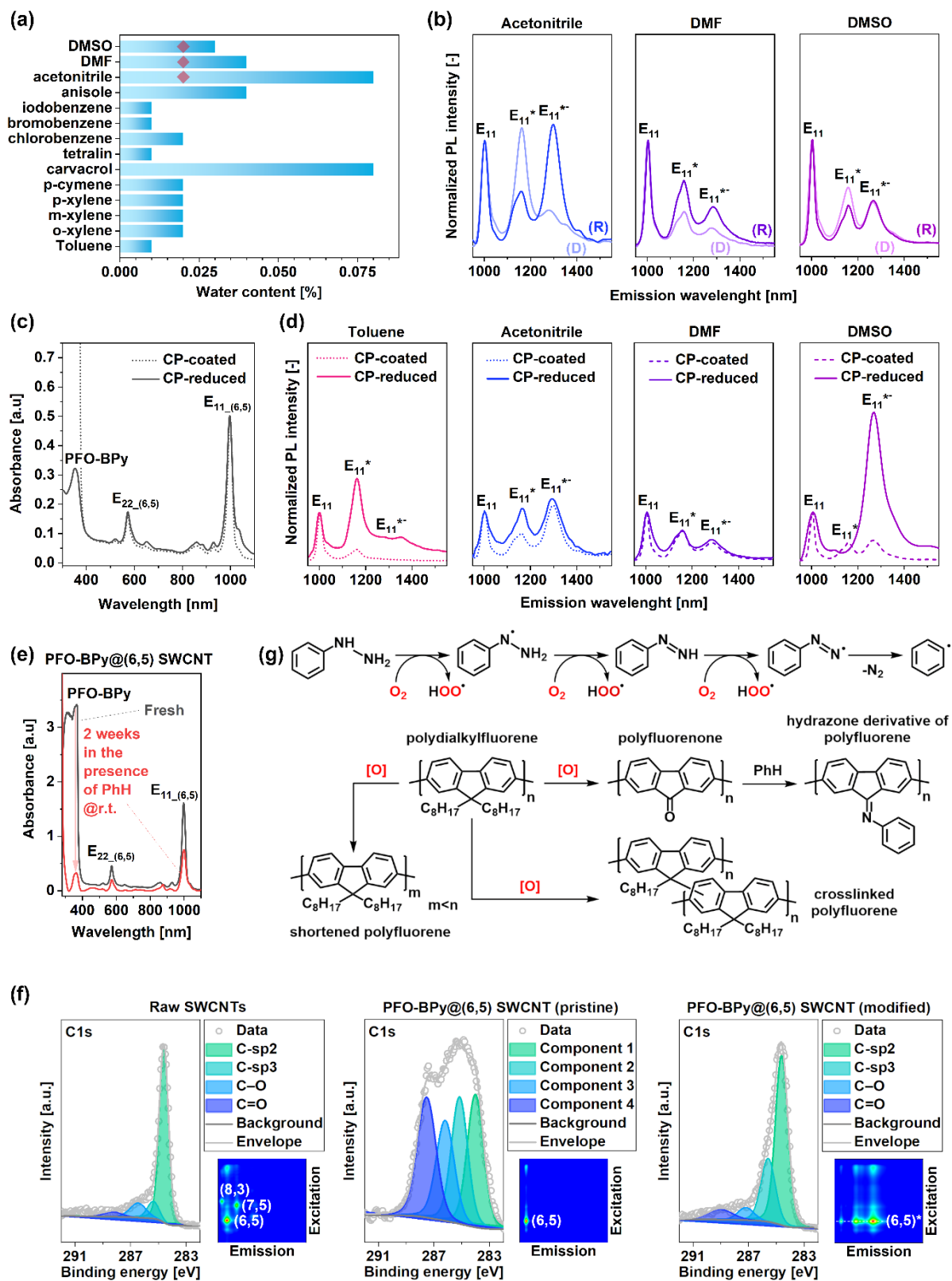

**Figure 3** (a) Moisture content in the solvents used for SWCNT functionalization (the content of water in selected solvents dried with molecular sieves examined in more detail in panel (b) is indicated with red rhombi), (b) PL spectra ($\lambda_{ex}$=579 nm) of (6,5) SWCNTs wrapped with PFO-BPy functionalized in the indicated solvents with PhH before (R – regular) and after drying (D – dried), (c) optical absorption spectra of (6,5) SWCNTs suspended with PFO-BPy in toluene before and after purifying from the polymer, and (d) PL spectra ($\lambda_{ex}$=579 nm) of (6,5) SWCNTs wrapped with PFO-BPy functionalized in the indicated solvents with PhH before and after removal of the conjugated polymer (CP) from the SWCNT surface, (e) optical absorption spectra of as-made PFO-BPy@(6,5) SWCNT suspension and after exposure to PhH for two weeks at room temperature, (f) C1s spectra recorded by XPS of the raw SWNCT powder, SWCNT suspension prepared by selective wrapping with PFO-BPy (components not defined), and PFO-BPy@(6,5) SWCNT subjected to functionalization with PhH (1 μl/mL) in DMSO at 80°C for 2 h. Insets show corresponding PL excitation-emission maps. (g) Reactions generating phenyl radicals from PhH and possible transformations experienced by the polyfluorene-based polymers and copolymers. Pyridine units are not shown for the sake of clarity.

We previously concluded that the Lewis base character of the solvent enhances the reactivity of SWCNT-based systems and promotes a higher degree of chemical modification, often resulting in the emergence of an $E_{11}^{*-}$ peak beyond 1,250 nm. Hence, the acidic character of the molecular sieves reduces this tendency and better supports the generation of an $E_{11}^{*}$ peak corresponding to moderately functionalized SWCNTs. The most striking difference, supporting this hypothesis, was observed in the case of acetonitrile, for which the preference for the formation of $E_{11}^{*}$ and $E_{11}^{*-}$ peaks flipped upon drying the solvent. The relatively high content of water, which can be considered as a Lewis base, in unprocessed acetonitrile, likely promoted the implantation of a substantial number of luminescent defects, compared to other solvents. It should also be highlighted that the position of the $E_{11}^{*-}$ peak maximum for acetonitrile reached 1,300 nm (Figure 2, Figure 3bd). This effect likely results from solvatochromism, to which the $E_{11}^{*-}$ peaks are more sensitive than $E_{11}^{*}$ and $E_{11}$ [35,55]. The spectrum of the PFO-BPy itself was most red-shifted in this solvent (Figure 3d), which strengthens this hypothesis. Thus, drying solvents using molecular sieves may not merely remove water but also induce crucial changes for the nanochemistry of low-dimensional materials.

Comparing the functionalization of polymer-free and polymer-wrapped SWCNTs, we see that polymer presence on the SWCNT surface may facilitate or hinder the introduction of luminescent defects into SWCNTs, depending on the solvent (Figure 1b, Figure 2b). At the same time, a complete lack of polymer favors the emergence of multiple $E_{11}^{*}$ features, especially those that originate from the radical transfer to the solvent ($E_{11\_(6,5)}^{*(AL)}$) at ca. 1,120 nm. On the contrary, the polymer-wrapped SWCNTs prefer aryl functionalization ($E_{11\_(6,5)}^{*(AR)}$), which manifests at ca. 1,160 nm. Therefore, we explored how reducing the polymer content in the selectively extracted (6,5) SWCNTs would affect their modification. Seeking the best materials for emission in the further parts of the NIR spectrum, we focused our analysis on SWCNTs processed in polar aprotic solvents, which displayed the most significant redshifts (Figures 1 and 2). Toluene was used as a reference solvent. In the literature, it is generally accepted that minimizing the polymer content improves the degree of SWCNT functionalization [22,39]. However, this knowledge is a generalization resulting from the evaluation of how SWCNTs behave in toluene. It is therefore desired to extend the scope of the liquid media to validate this relatively widespread assumption. The optical absorption spectra demonstrated successful reduction of the amount of polymer, the intensity of the optical transition of which decreased by two orders of magnitude (Figure 3c) What is interesting, while the reduction of the PFO-BPy content on the surface of (6,5) SWCNTs did indeed improve the reactivity of SWCNTs in toluene (Figure 3d), the effect of such processing was more complex for other solvents. In the case of acetonitrile, the relative intensity of $E_{11}$* and $E_{11}$*- peaks increased (the former optical feature in particular). Acetonitrile may have enlarged the solvent-accessible surface area by rearranging the polymer molecules on the SWCNT surface, thereby reducing the likelihood that two functional groups will attach in the vicinity of one another. Furthermore, the functionalization of (6,5) SWCNTs with reduced polymer content in DMF using PhH also improved the functionalization degree, but to a smaller extent. Finally, the most beneficial result was obtained for the covalent modification of SWCNTs in DMSO after partial removal of the polymer. Under such conditions, the SWCNTs were substantially modified, and almost exclusive $E_{11\_(6,5)}$*-$^{(AR)}$ defect emission centered at 1,280 nm was noted.

Hence, the presence of the polymer significantly influences the modification process. Such polymer coating not only affects the degree of chemical modification reactivity, but the emergence of different optical spectra strongly suggests that the polymer molecules may impact the type and concentration of reactive species in the system. Interestingly, the polymer itself is not resistant to the reactive species present in the medium and displays a self-cleaning capacity. Such a feature has not been reported for other SWCNT modification strategies, making this functionalization platform unique. Removal of the polymer from the SWCNT surface is highly challenging [65], which results in considerable material loss. Hence, efforts have been made to design depolymerizable dispersants [66–68]. Unfortunately, such conjugated polymers do not exhibit similarly impressive SWCNT extraction selectivity as the well-known PFO-BPy [69], PFO-BT [53], or PFO [70]. The recorded optical absorption spectra show that upon leaving the PFO-BPy@(6,5) SWCNT sample exposed to PhH for a prolonged period, the intensity of the peak corresponding to the polymer was dramatically reduced (Figure 3e). To validate that the treatment of SWCNTs with PhH may favorably disturb the polymer corona around SWCNTs, we examined the material by XPS (Figure 3f). The raw SWCNT powder was primarily made of $sp^2$-hybridized carbon atoms, judging by the sharp nature of the C1s peak. Upon coating of the material with PFO-BPy, considerable C1s peak broadening was observed. The data resembled a typical spectrum of a conjugated polymer [71], which matched the expectations since PFO-BPy creates a dense layer around SWCNTs [54]. Finally, upon treating the PFO-BPy@(6,5) SWCNT material with PhH, the resulting C1s peak was not broad. Compared with the spectrum registered for untreated SWCNTs, we noted an increased content of $sp^2$-hybridized carbon caused by addition of luminescent defects. Hence, we proved that exposing SWCNTs to PhH alleviated the issue of excessive unwanted polymer molecules on SWCNTs.

Several chemical transformations may be responsible for this effect (Figure 3g). Firstly, the conversion of PhH to phenyl radicals generates hydroperoxyl radicals (HOO•) [39], which are potent oxidants [72]. In the presence of these and other reactive species, as well as oxygen dissolved in toluene, the polyfluorene framework may be subject to main chain scission. The increased length of the side chain promotes such a breakup, so polymers made of dioctylfluorene units may undergo this undesired transformation, reducing their molecular weight [73] and increasing the probability of their detachment from the SWCNT surface [54]. Secondly, a common problem associated with the application of polyfluorene compounds is caused by their tendency to undergo oxidation to a corresponding ketone, i.e., polyfluorenone [74–77]. This degradation mechanism, which hinders the use of polyfluorene-based light-emitting devices, is likely to occur under the above-explained oxidizing conditions. As shown in the cited literature, this issue is also persistent in dialkyl-substituted fluorenes. In the case described herein, the situation is more complex as PhH may react with the as-formed carbonyl group of the polyfluorenone to produce a corresponding hydrazone derivative [78]. Thirdly, in the presence of oxygen and reagents of similar nature, polyfluorene may exhibit crosslinking [79,80]. Considering that the above-described results show that uncovering the SWCNT surface increases the reactivity of SWCNTs, the capacity of the PhH-driven functionalization system to perform polymer rearrangement autonomously is highly beneficial.

### 2.3. Multi-scale modelling of SWCNT-polymer-solvent interactions

To gain deep insight into the impact of various solvents on the interaction between polymers and (6,5) SWCNTs, we employed spin-polarized density functional theory (DFT) [81–84], incorporating solvent effects via the COSMO continuum solvation model [85,86], as well as molecular dynamics (MD), time-stamped force-bias Monte Carlo (MC)[87,88], and density functional-based tight-binding (DFTB) modelling[89,90] (see Section 4 in the SI for details). Given that charge transfer between a conjugated polymer and a specific SWCNT type is a key factor influencing chirality-selective extraction[91], we analyzed the energy level diagram, aligning the calculated HOMO and LUMO levels of monomeric, trimeric, pentameric, and infinite-chain PFO-BPy6,6' polymers to the projected density of states (PDOS) of the SWCNT (Figure 4). To ensure the robustness of these calculations, we compared results obtained using a newly implemented HSE06 exchange-correlation functional [53,92] in SIESTA [93,94] with those from QuantumATK [95,96] (Figure S4). HOMO and LUMO spatial distributions are consistent across both approaches, confirming that the observed electronic trends are independent of basis set and code-specific implementations. To assess the impact of solvent environments on the polymer's frontier orbitals, we performed calculations both in vacuum and using the COSMO model to simulate toluene, acetonitrile, and DMSO environments.

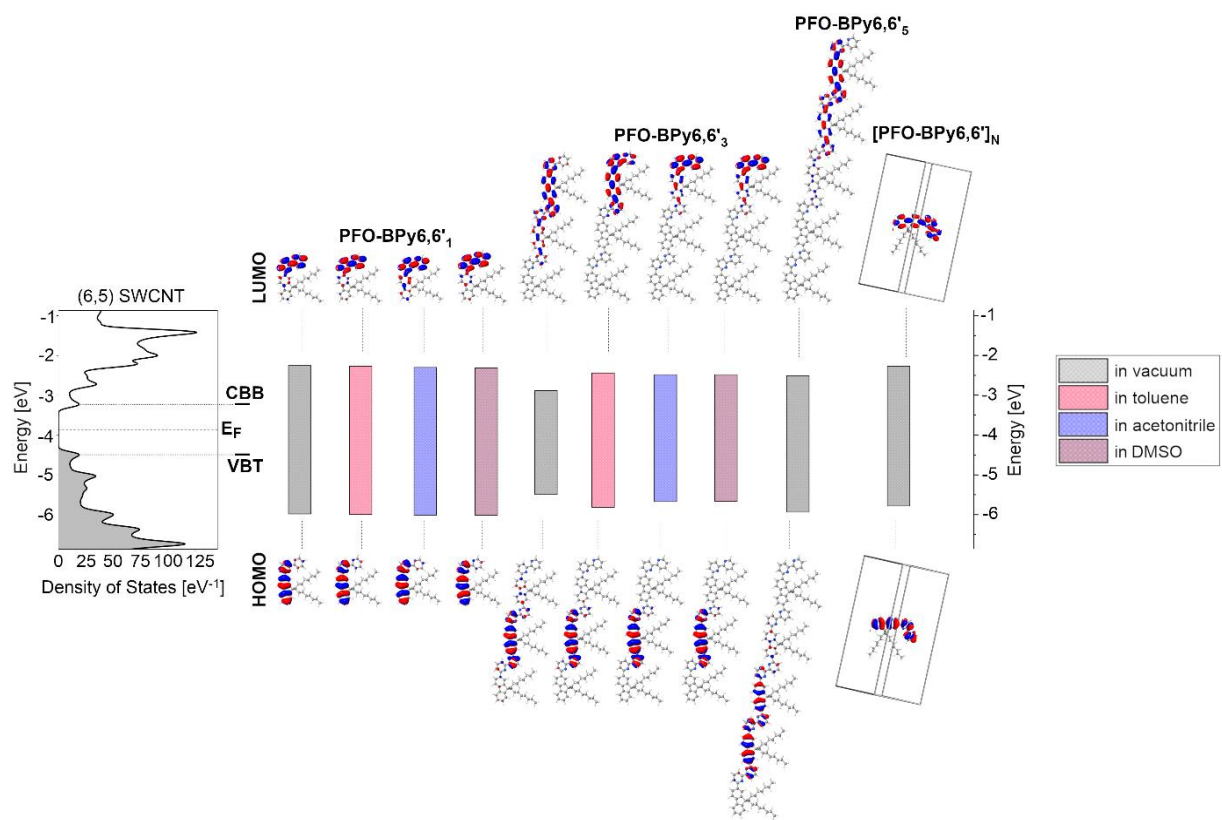

**Figure 4** Projected density of states of (6,5) SWCNT and energy levels of PFO-BPy6,6' polymer of varying lengths calculated using spin-polarized density functional theory (SP-DFT) with the HSE06 hybrid functional, PseudoDojo pseudopotentials, and a medium basis set. Polymer energy levels are aligned to the Fermi level ($E_F$) of the SWCNT and presented in absolute energy units. The valence band top (VBT) and conduction band bottom (CBB) of the SWCNT are indicated. HOMO and LUMO levels of fully relaxed polymers are depicted as vertical bars, calculated in four different environments: vacuum (grey), toluene (pink), acetonitrile (navy), and DMSO (purple). The spatial distributions of the HOMO and LUMO orbitals are visualized at an isovalue of 0.06 Å$^{-3/2}$, with red and blue lobes representing the negative and positive regions of the wavefunction, respectively. On the far right, the orbital distribution for the infinite polymer is shown within a simulation cell. Carbon, nitrogen, and hydrogen atoms are colored gray, blue, and white, respectively.

For the monomer, HOMO and LUMO energy levels remained essentially congruous across different environments. However, for the trimer, the presence of solvent led to a stabilization of the HOMO levels (more negative values) and a destabilization of the LUMO levels (less negative values), effectively narrowing the HOMO–LUMO gap. This trend aligns with literature reports indicating that solvent polarity can influence the electronic properties of conjugated systems [97,98]. Regarding orbital shapes, HOMOs were relatively delocalized across the polymer backbone in all environments. In contrast, LUMOs exhibited increased localization on the bipyridine (BPy) units with chain elongation, a phenomenon that was more pronounced in polar solvents such as acetonitrile and DMSO. This enhanced localization suggests stronger electron-accepting behavior of the BPy units in polar environments. For the infinite polymer chain, calculations performed in vacuum revealed that both the HOMO and LUMO orbitals were delocalized along the polymer backbone, indicating a uniform electronic distribution. The energy level alignment suggests that the polymer's HOMO lies below the SWCNT valence band edge, and its LUMO lies above the conduction band edge. This configuration corresponds to a broken-gap or type-Ib heterojunction, suggesting that direct charge transfer under equilibrium conditions may be energetically unfavorable. However, photoinduced or field-assisted processes could still enable charge separation, depending on the environment and polymer length.

Closer inspection of polymer morphologies reveals substantial solvent-induced structural reorganization in PFO-BPy6,6' polymers, highlighting that vacuum-optimized geometries do not adequately capture conformational behavior under realistic conditions. The extent of these structural changes correlates with solvent polarity and is most pronounced in polar environments such as acetonitrile and DMSO, where deviations from vacuum structures are the greatest (Figure S5). Overall, structural reorganization is more significant for the trimer than for the monomer in any given solvent, likely due to the increased conformational flexibility and the larger number of solvent-sensitive sites along the extended backbone and side chains. As expected, the most pronounced deviations occur at the polymer termini, involving not only the bipyridine (BPy) units but also the alkyl side chains, suggesting that both the conjugated core and peripheral groups respond to the surrounding dielectric environment.

These solvent-induced structural reorganizations are accompanied by pronounced changes in solvation energetics (Figure S6). While the spatial distribution of the solvent-accessible surface charge density remains broadly consonant across solvents, the total solvation energy varies significantly with both solvent polarity and polymer length. This result highlights that even subtle geometric rearrangements can lead to substantial differences in electrostatic stabilization. The most negative surface charge densities are mainly located on the alkyl side chains. At the same time, the nitrogen atoms in the conjugated backbone exhibit the most positive regions, indicating that both polar and dispersive interactions play a role in modulating solvation.

We performed additional DFTB simulations to identify the influence of solvent on polymer-SWCNT interactions. Due to computational limitations, the systems used for DFTB modeling were constructed by reducing the size of those used in the MD/MC simulations shown in Figure S7. Specifically, smaller subsystems were extracted from the final MD/MC snapshots, preserving a ~8 Å thick solvent shell around both the SWCNT and the polymer. Cross-sectional views obtained from the DFTB simulations (Figure 5a) reveal pronounced solvent-dependent differences in how the PFO-BPy$_{6,6'}$ polymer conforms to the (6,5) SWCNT, highlighting structural features not captured in classical MD/MC simulations. In acetonitrile and DMSO, the polymer wraps more closely around the SWCNT, with a greater number of atoms located in close proximity to the SWCNT surface. At the same time, portions of the polymer—particularly the side chains—extend further away from the SWCNT compared to the toluene case. This suggests that in polar solvents the polymer adopts a more spatially extended but closely wrapping conformation. In contrast, in toluene, the polymer appears more condensed overall, yet less intimately wrapped around the SWCNT. These structural differences are further illustrated in Figure S9, where a different viewing angle of the coloured-atom visualization shows how polymer atoms are distributed along the SWCNT, helping to identify which regions of the polymer lie closer or further from the SWCNT surface.

These structural variations are accompanied by substantial differences in electrostatic potential (Figure 5b) and charge distribution (Figure 5c) across the three solvent systems. A comparison of the electrostatic difference potential ($\Delta V_E$) along the SWCNT interacting with PFO-BPy6,6' showed a marked contrast between the system in toluene and those in acetonitrile and DMSO. $\Delta V_E$ highlights spatial regions where significant charge redistribution occurs, relative to the sum of electron densities of neutral, non-interacting atoms (Figure 5b). Both acetonitrile and DMSO induce much larger variations in $\Delta V_E$ than toluene. In all cases, $\Delta V_E$ oscillates around zero in regions where the SWCNT interacts with the polymer. Immersing the SWCNT–polymer complex in acetonitrile results in more positive $\Delta V_E$ values when the polymer side chains are adjacent to the SWCNT, and more negative values when the polymer backbone is nearby. In contrast, the system in DMSO exhibits an opposite trend: $\Delta V_E$ becomes more negative when the side chains are near the SWCNT and more positive when the backbone is closer.

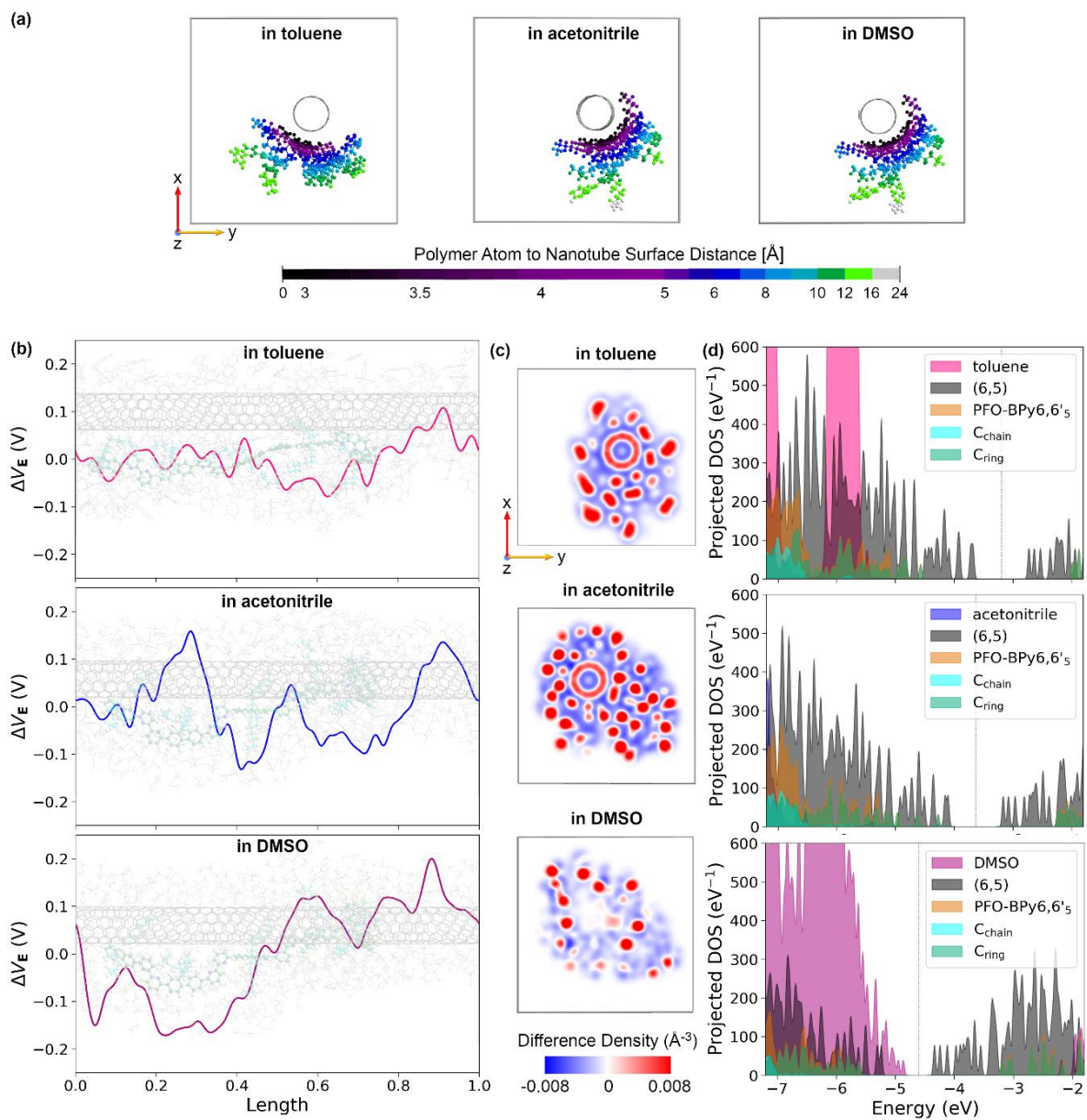

**Figure 5** Computed structural and electronic properties of (6,5) SWCNT interacting with PFO-BPy6,6'$_5$ polymer in toluene, acetonitrile, or DMSO solutions obtained at the DFTB level. (a) Atomistic cross-sectional views of each system. SWCNTs and polymers are shown using stick and ball-and-stick models, respectively. Solvent molecules are omitted for clarity. Polymer atoms are colored based on their distance from the SWCNT lateral surface, with darker shades indicating closer proximity and lighter shades indicating greater distance. (b) Electrostatic difference potential ($\Delta V_E$) plotted as a 1D projection on the z-axis (along SWCNT symmetry axis). $\Delta V_E$ is the difference between the electrostatic potential of the self-consistent valence charge density and the electrostatic potential from a superposition of atomic valence densities. SWCNTs and polymers represented by stick and ball-and-stick models are shown on all cut-planes, while solvent molecules are represented as lines. (b) Electron difference density (EDD) maps showing the difference between the self-consistent valence charge density and a superposition of atomic valence densities. The XY cut planes are taken at 0.4 (across the SWCNT and polymer complex) of the fractional length of the simulation boxes along the z-axes. (c) The projected density of states (PDOS), shown on an absolute energy scale, is presented for the SWCNT, polymers, solvents, and distinct regions ($C_{ring}$ and $C_{chain}$) of the polymers for both systems. The Fermi level is indicated by a dotted line.

These solvent-dependent differences suggest that the nature of the polymer–SWCNT interaction is strongly modulated by the surrounding dielectric environment. This effect is also clearly visible in the cross-sectional charge density maps shown in Figure 5c. In both toluene and acetonitrile, a blue region—indicating electron deficiency—is observed inside the SWCNT, while a red region—representing electron accumulation—is localized on the SWCNT wall. This redistribution pattern is absent in DMSO, where the difference density across the SWCNT cross-section is nearly zero. Solvent-dependent variations in electrostatic potential and charge density reveal that polar solvents such as acetonitrile and DMSO significantly modulate the electronic environment of the polymer–SWCNT interface. While toluene preserves weak, localized interactions, polar solvents induce stronger or inverted charge redistribution patterns, suggesting solvent-mediated control over polymer–SWCNT coupling.

Solvent-mediated differences in the interaction between the PFO-BPy6,6′ and (6,5) SWCNT are also reflected in the electronic structure of the combined systems. Figure 5c presents the projected density of states (PDOS) for the solvent, SWCNT, polymer, and distinct polymer regions (main chain/side chains) in each solvent environment. A shift of the Fermi level toward more negative energies is observed in the polymer–SWCNT complex immersed in acetonitrile, compared to toluene. An even larger shift to lower energies occurs in the presence of DMSO. All states associated with the SWCNT and polymer in the (6,5)+PFO-BPy6,6′+DMSO system are shifted to lower absolute energy values relative to those in the toluene-based system. Notably, in contrast to the toluene and acetonitrile systems, DMSO-related states appear within the band gap region, effectively reducing the system's band gap and forming a new valence band edge. Additionally, the DMSO states show greater overlap with both SWCNT and polymer states over a broader energy range than is observed for toluene or acetonitrile, suggesting stronger electronic coupling and more pronounced solvent–system hybridization.

### 2.4. The application of chemically modified SWCNTs for cholesterol sensing

Currently, SWCNT-based optical sensors are widely used for detecting biogenic amines, such as dopamine and serotonin [8,32,89]. However, this low-dimensional material can sense other vital molecules. Cardiovascular diseases are the leading cause of death globally [99]. A reduction in cholesterol concentration in the blood could considerably decrease the risk of death due to this condition [100], so it would be valuable to have the capacity to track the amount of this compound in a more accessible manner. Recently, SWCNTs enhanced with luminescent defects have revealed that the optical characteristics of such materials are sensitive to cholesterol [10].

However, with the increase in cholesterol concentration, the luminescence intensity of the SWCNTs generally decreased. This effect reduces the utility of such nanosensors, as the higher the cholesterol concentration, the more challenging it becomes to detect the optical signature of its presence.

Regardless of cholesterol concentration, the PL spectra of pristine post-CPE SWCNTs remained intact, as the absolute intensity of the $E_{11}$ peak remained unchanged (Figure 6a). However, the SWCNTs (covalently modified according to the methodology reported herein), which displayed the defect-related $E_{11}$* (Figure 6b) and $E_{11}$*- (Figure 6c) peaks created in toluene and DMSO, respectively, exhibited a different behavior. In both cases, upon exposure to cholesterol, the fluorescence intensity of such SWCNTs increased. A gradual increase in cholesterol concentration resulted in a corresponding brightening of SWCNTs, facilitating cholesterol sensing (Figure 6d). This outcome highlights the high application potential of the materials designed in this study, whose optical characteristics were tuned by selecting an appropriate solvent for functionalization.

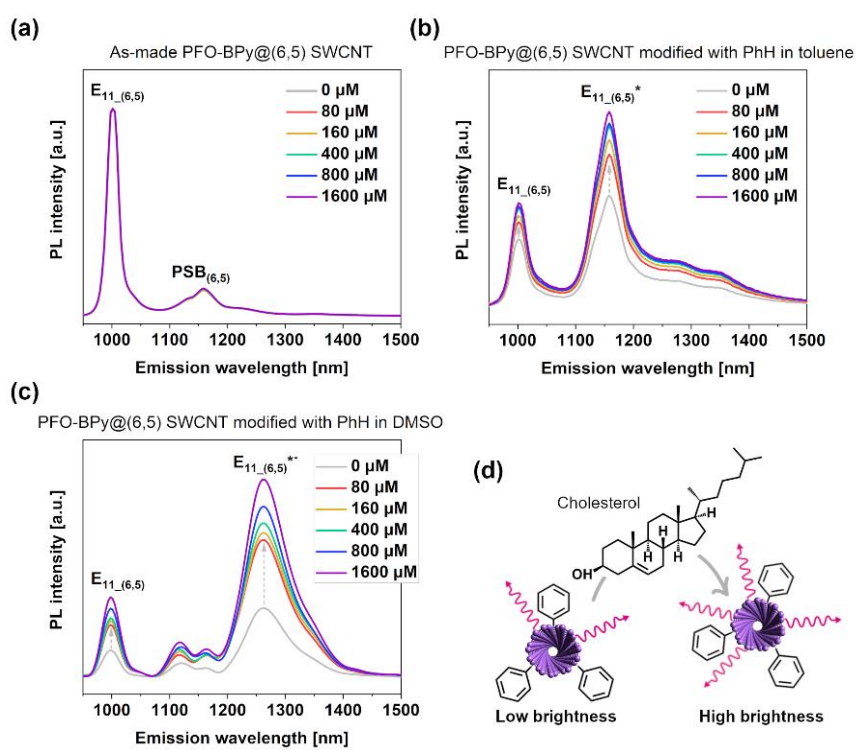

**Figure 6** PL spectra ($\lambda_{ex}$=579 nm) of (a) pristine (6,5) SWCNTs wrapped with PFO-BPy and functionalized with PhH in (b) toluene, and (c) DMSO as a function of concentration of cholesterol, to which the samples were exposed. (d) Brightening of SWCNTs modified with luminescent defects in the presence of cholesterol. PSB – photoluminescence sideband [22].

## Conclusions

The carried-out work sheds considerable insight into the solvent-dispersant-SWCNT interactions, which determine the reactivity of SWCNTs. The inclusion of polymer molecules on the SWCNT surface significantly impacts the types of radicals generated and subsequently attached to the SWCNTs, which in turn determine their resulting optical properties. This finding clarifies a common misconception that the dispersant molecules merely block the access of reactive species to SWCNTs. Interestingly, we also discovered that polar solvents promote high-degree functionalization of SWCNTs, which is manifested by the emergence of red-shifted optical features coinciding with the telecommunication part of the near-infrared range. The very much improved propensity of SWCNTs for functionalization in polar solvents results from better charge redistribution at the polymer–SWCNT interface occurring in such chemical environments. Hence, this newly noted relationship provides powerful means to tailor the surface of SWCNTs and other relevant low-dimensional materials. Finally, the SWCNTs crafted according to the reported strategy have the capacity to detect cholesterol, and as cholesterol levels increase, their brightness improves, further enhancing their utility.

## Conflicts of interest

There are no conflicts of interest to declare.

## Acknowledgments

D.J., R.S., and D.J. would like to thank the National Science Centre, Poland (under the SONATA program, Grant agreement UMO-2020/39/D/ST5/00285) for supporting the research. They also appreciate the Silesian University of Technology Rector's pro-quality grant for groundbreaking research (32/014/SDU/10-21-20). K.Z.M. and O.E.-R. gratefully acknowledges the Interdisciplinary Centre for Mathematical and Computational Modelling at the University of Warsaw, Poland (Grant No. G47-5), and DIPC Supercomputing Center, Spain, for providing computer facilities and technical support. Computing facilities were also kindly provided by the Theory of Condensed Matter Group at the Cavendish Laboratory, University of Cambridge, UK. K.Z.M., O.E.-R., and Y.P. are grateful to Agencia Estatal de Investigación, Ministerio de Ciencia e Innovación, Spain (Proyectos de Generación de Conocimiento 2022 program, PID2022-139776NB-C65) for funding this research. K.Z.M. also would like to thank the European Commission (Marie Skłodowska-Curie Cofund Programme; grant no. H2020-MSCA-COFUND-2020-101034228-WOLFRAM2) for the financial support of the study. J.J. acknowledges financial support from Grant No. PID2022-139776NB-C63 funded by MCIN/AEI/10.13039/501100011033 and by ERDF/EU "A way of making Europe" by the European Union.

Supplementary Information

# Polar Express: Rapid Functionalization of Single-Walled Carbon Nanotubes in High Dipole Moment Media

Dominik Just[1,*], Ryszard Siedlecki[1], Maciej Krzywiecki[2], Oussama Er-Riyahi[3], Yann Pouillon[3], Javier Junquera[4], Karolina Z. Milowska[3,5], Dawid Janas[1,*]

[1] Department of Chemistry, Silesian University of Technology, B. Krzywoustego 4, 44-100, Gliwice, Poland

[2] Institute of Physics – Centre for Science and Education, Silesian University of Technology, Konarskiego 22B, 44-100, Gliwice, Poland

[3] CIC nanoGUNE, Donostia-San Sebastián 20018, Spain

[4] Departamento de Ciencias de la Tierra y Física de la Materia Condensada, Universidad de Cantabria, Avenida de los Castros, s/n, E-39005 Santander, Spain

[5] Ikerbasque, Basque Foundation for Science, Bilbao 48013, Spain

* Corresponding author(s): dominik.just@polsl.pl, dawid.janas@polsl.pl

# Table of Contents

## 1. Materials

The materials used in the experiments were employed in the form in which they were supplied. The purity of the reagents, along with the manufacturer and data allowing for identification, can be found below.

### 1.1. Raw SWCNT materials

This study was carried out using (6,5)-enriched CoMoCAT SWCNTs (Sigma Aldrich, product number: 773735, lot: MKCM5514, purity: 95%-carbon basis).

### 1.2. Reagents for CP synthesis

9,9-dioctylfluorene-2,7-bis(boronic acid pinacol ester) (Angene, cat. number: AG0034EZ, CAS: 196207-58-6, purity: 98%), 2,7-dibromo-9,9-dioctyl-9H-fluorene (Sigma Aldrich, cat. number: 560073-25g, CAS: 198964-46-4, purity: 96%), Aliquat 336 TG (Alfa Aesar, cat. number: A17247, CAS: 63393-96-4, purity: N/A), tetrakis(triphenylphosphine)palladium – $Pd(PPh_3)_4$, (Apollo Scientific, cat. number: OR4225, CAS: 14221-01-3, purity: >99%).

### 1.3. Solvents for CP synthesis and SWCNT sorting/functionalization

Toluene (Alfa Aesar, cat. number: 19376.K2, CAS: 108-88-3, spectrophotometric grade, purity: >99.7%), dimethylformamide (Acros Organics, cat. number: 348435000, CAS: 68-12-2, purity: 99.8%, extra dry over molecular sieves AcroSeal®), methanol (Chempur, cat. number: 116219904, CAS: 67-56-1, purity: pure p.a.), ethanol (Stanlab, cat. number: 603-002-00-5, CAS: 64-17-5, purity: 96%), chloroform (Stanlab, cat. number: 062-006-00-4, CAS: 67-66-3, purity: p.a.), acetone (ChemLand, cat. number: 102480111, CAS: 67-64-1, purity: pure p.a.), iodobenzene (Alfa Aesar, cat. number: A12788, CAS: 591-50-4, purity: 98%), bromobenzene (Acros Organics, cat. number: A0417929, CAS: 108-86-1, purity: 99%), p-cymene (Thermo Scientific, cat. number: A19226.AP, CAS: 99-87-6, purity: 97%), benzene (Chempur, cat. number: 111625000, CAS: 71-43-2, purity: purity p.a.), chlorobenzene (Thermo Scientific, cat. number: 2052185, CAS: 108-90-7, purity: ≥99%), carvacrol (Ambeed, cat. number: A288147, CAS: 499-75-2, purity: 99.92%), o-xylene (Thermo Scientific, cat. number: A11358, CAS: 95-47-6, purity: 99%), m-xylene (Thermo Scientific, cat. number: L03788, CAS: 108-38-3, purity: 99%), p-xylene (Thermo Scientific, cat. number: A10534, CAS: 106-42-3, purity: 99%), acetonitrile (Thermo Scientific, cat. number: A/0620/PB08, CAS: 75-05-8, purity: ≥99%), 1,2,3,4-tetrahydronaphtalene (Thermo Scientific, cat. number:

1467, CAS: 119-64-2, purity: ≥98%), dimethyl sulfoxide (Chempur, cat. number: 113635510, CAS: 67-68-5, purity: HPLC)

### 1.4. Reagents for SWCNT functionalization

Phenylhydrazine (Alfa Aesar, cat. number: A11246. 18, CAS: 100-63-0, purity: 97%), aniline (Chempur, cat. number: chem*111458308*250ml, CAS: 62-53-3, purity: p.a.), and cholesterol (Thermo Scientific, cat. number: 110190250, CAS: 57-88-5, purity: 95%)

## 2. Methods

### 2.1. Polymer synthesis

The syntheses of PFO and PFO-BPy (Table S1) were carried out in accordance with the Suzuki coupling procedure presented below. The structures of the obtained polymers were confirmed using $^1$H NMR spectroscopy, and their macromolecular parameters were determined with size exclusion chromatography (SEC). In brief, organoboron derivative PA (9,9-di-n-alkylfluorene-2,7-diboronic acid bis(pinacol)ester, purity: 98%) (0.500 g, 0.763 mmol, 1.00 mol. eq.) and a sufficient amount of dibromo derivative PB (9,9-dialkyl-2,7-dibromofluorene purity: 96%) (0.418 g, 0.763 mmol, 1.00 mol. eq.) were added to the high-pressure glass reaction vessel. The reactor was then filled with 1M $Na_2CO_3$ solution (12 mL) and toluene (12 mL). Three drops of Aliquat 336 phase transfer catalyst (PTC) were added. The mixture was purged with argon for 30 min, and subsequently, $Pd(PPh_3)_4$ (0.022 g, 0.019 mmol, 0.025 eq.) was added. The reaction mixture was stirred vigorously at 80°C for 4 days. Afterward, the reaction mixture was cooled and poured into a mixture of methanol and water in a 6:1 ratio. The primary product was then filtered off, dissolved in chloroform, and dried using $MgSO_4$. The collected material was evaporated to dryness and dissolved in a sufficient volume of chloroform. The final product was precipitated in methanol. The fibrous polymer was collected by filtration, washed twice with 50 mL of cold methanol, and finally twice with 50 mL of cold acetone.

**Table S1** List of obtained CPs with the various subunits used for the synthesis.

| Name | Acronym | PA | PB | CP no. |
|---|---|---|---|---|
| Poly(9,9'-dioctylfluorenyl-2,7-diylalt-6,6'-(2,2'-bipyridine)) | PFO-BPy | 9,9-di-n-octylfluorene-2,7-diboronic acid | 6,6'-dibromo-2,2'bipyridine | P1 |
| Poly(9,9'-dioctylfluorenyl-2,7-diyl) | PFO | | 9,9-dioctyl-2,7-dibromofluorene | P2 |

## 2.2. CPE process

3 mg of SWCNTs and 12 mg of CP were introduced to a 19 mL glass vial. 12 mL of toluene was then added to the mixture and homogenized in a bath sonicator for 15 minutes (Polsonic, Sonic-2, 250 W) at 5°C. Further, more vigorous sonication with a tip sonotrode (Hielscher UP200St ultrasonic generator) was performed to disentangle SWCNTs and wrap them with the CP molecules for 16 minutes at a power of 30 W. After sonication, the obtained suspension was transferred to a 15 mL conical tube and centrifuged at 10,000 rpm (15,314 × g) for 4 minutes to precipitate the bundled and non-wrapped SWCNTs, as well as CP aggregates. The lump-free generated supernatant was transferred to a fresh vial and subjected to analysis by UV-Vis spectroscopy and PL excitation-emission mapping.

## 2.3. Functionalization of raw SWCNTs in various solvents

1.5 mg of raw (6,5)-enriched SWCNTs raw (CoMoCAT, SG65i) was dispersed in 5 mL of each of the indicated solvents (Table S2) using a tip-horn sonicator, then 2.75 mg of PhH was added to the dispersion, vortexed, and then heated for one hour at 80°C. After thermal functionalization, (2,2,6,6-Tetramethylpiperidin-1-yl)oxyl (TEMPO) was added to the material to trap radicals resulting from PhH decomposition, bath sonicated, vortexed, and filtered using a Teflon membrane. Consequently, the radicals were quenched, and non-SWCNT compounds were removed. For characterization purposes, purified functionalized (6,5) SWCNTs were selectively extracted PFO-BPy6,6' in toluene and then analyzed by PL spectroscopy. To avoid detector saturation and to ensure the reliability of the measurements, the SWCNT concentration was set to an absorbance value of 0.5 a.u. before the measurements.

**Table S2** List of solvents with their physical properties used in SWCNT modification

| Solvent | Dipole Moment [D] | Density [g/mL] | Dielectric Constant [F/m] | Boiling Point [°C] | Viscosity at 20°C [mPa·s] |
|---|---|---|---|---|---|
| Benzene | 0.00 | 0.876 | 2.274 | 80.1 | 0.61 |
| Toluene | 0.36 | 0.867 | 2.38 | 110.6 | 0.59 |
| *o*-xylene | 0.62 | 0.880 | 2.56 | 144.4 | 0.80 |
| *m*-xylene | 0.37 | 0.864 | 2.31 | 139.1 | 0.64 |
| *p*-xylene | 0.00 | 0.861 | 2.27 | 138.3 | 0.60 |
| *p*-cymene | 0.00 | 0.857 | 2.72 | 177.0 | 0.85 |
| Carvacrol | 1.7 | 0.9772 | 9.8 | 237.7 | 6.1 |
| Tetralin | 0.61 | 0.970 | 2.32 | 207.0 | 2.35 |
| Chlorobenzene | 1.54 | 1.106 | 5.62 | 132.0 | 0.80 |
| Anisole | 1.38 | 0.995 | 4.33 | 154.0 | 1.30 |
| Acetonitrile | 3.92 | 0.786 | 35.9 | 81.6 | 0.36 |
| Dimethylformamide | 3.82 | 0.944 | 36.7 | 153.0 | 0.80 |
| Dimethyl sulfoxide | 3.96 | 1.100 | 47.2 | 189.0 | 2.00 |
| Water | 1.85 | 0.998 | 78.5 | 100.0 | 1.00 |

## 2.4. Functionalization of selectively-extracted SWCNTs in various solvents

1.5 mg of raw (6,5)-enriched SWCNTs (CoMoCAT, SG65i) was dispersed in the presence of 6 mg of PFO-BPy or 9 mg of PFO in toluene using a tip-horn sonicator to obtain near monochiral (6,5) and (7,5) SWCNT fractions. Then, toluene was evaporated, and the remaining precipitate was redispersed in a spectrum of indicated solvents using a sonication bath. Subsequently, the material was subjected to covalent modification following the methodology described in Section 2.5. Once the 1-hour-long heating was complete, TEMPO was added to halt the progress of the reaction. The mixture was then filtered and redispersed in 1 mg/mL toluene solutions of PFO-BPy and PFO. The process of transferring SWCNTs back to toluene was crucial for performing the measurement, because it is not possible to maintain them in a stable state in some of the liquid media. The concentration of the SWCNTs was adjusted to 0.5 a.u. (measured at the $E_{11}$ optical transition) for suspensions prepared in various solvents.

## 2.5. Determination of water content via Karl Fischer titration

To precisely quantify the residual water present in the solvents used for SWCNT functionalization, we employed Karl Fischer titration using a Metrohm 915 KF Ti-Touch instrument. This method, widely regarded for its high sensitivity and selectivity toward water, was chosen to accurately assess how trace moisture influences the outcome of chemical modifications. A series of measurements was performed on untreated (non-dried) solvents as well as on representative solvents that were dried using 4 Å molecular sieves for 2 weeks. The comparative analysis enabled a direct evaluation of the effectiveness of the drying procedure.

## 2.6. CP removal from the SWCNT surface

In order to wash off excess CP molecules from the SWCNTs, the SWCNT dispersions were first precipitated from the solution using methanol. Then, the material was purified by membrane filtration (Ahlstrom, PTFE, pore size: 2 µm, membrane thickness: 40µm ± 5µm) with hot toluene (80°C). Then, the solid was washed twice with methanol and chloroform, in which the CP dissolves very well. Finally, the membrane was washed with toluene to remove the previously used solvents. The SWCNTs prepared in this way were then redispersed in toluene using a different polymer for analysis of its impact on the reactivity of SWCNTs.

## 3. Characterization

### 3.1. Nuclear Magnetic Resonance ($^1$H NMR)

Proton nuclear magnetic resonance ($^1$H NMR) spectra of CPs (Figures S1–S2) were registered using a Varian Unity Inova spectrometer operating at 400 MHz. $^1$H-chemical shifts were measured in δ (ppm), using the characteristic peak for $CDCl_3$ (residual peak δ = 7.26 ppm) as a reference.

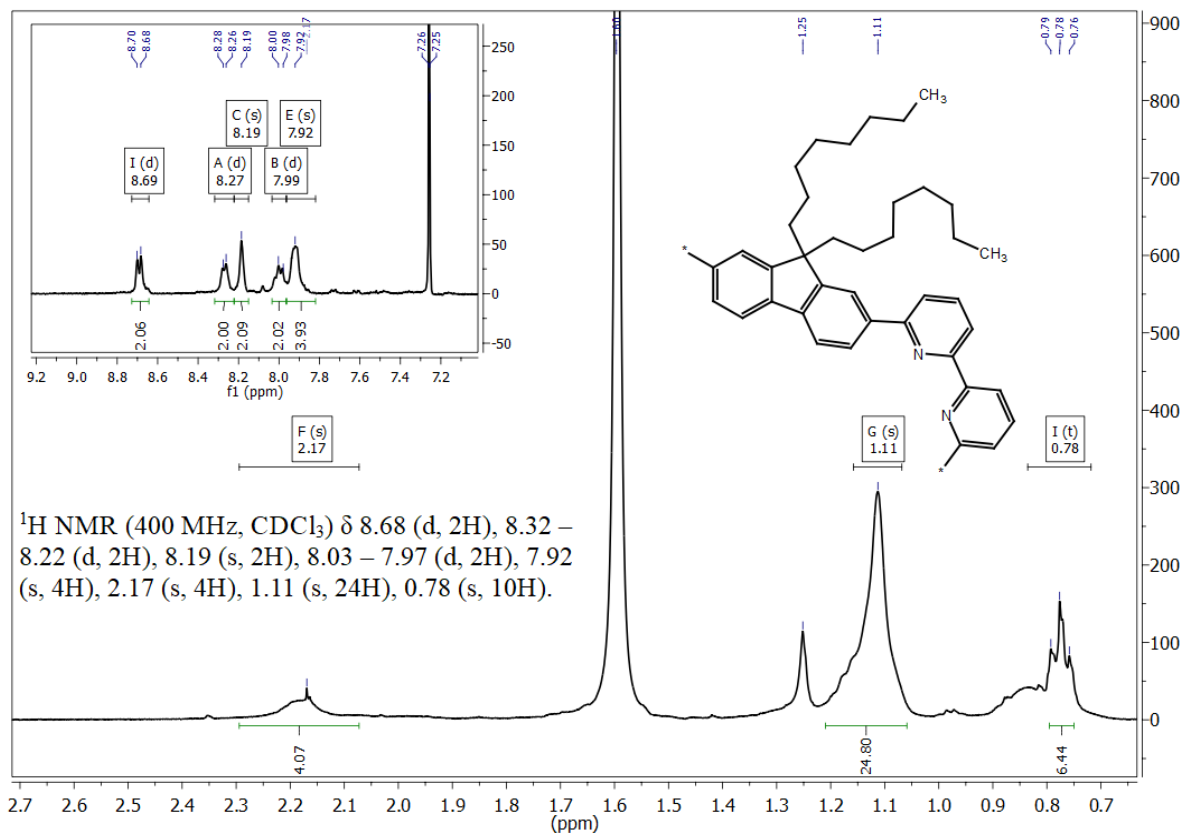

¹H NMR (400 MHz, CDCl₃) δ 8.68 (d, 2H), 8.32 – 8.22 (d, 2H), 8.19 (s, 2H), 8.03 – 7.97 (d, 2H), 7.92 (s, 4H), 2.17 (s, 4H), 1.11 (s, 24H), 0.78 (s, 10H).

**Figure S1** ¹H NMR spectrum of PFO-BPy synthesized in-house.

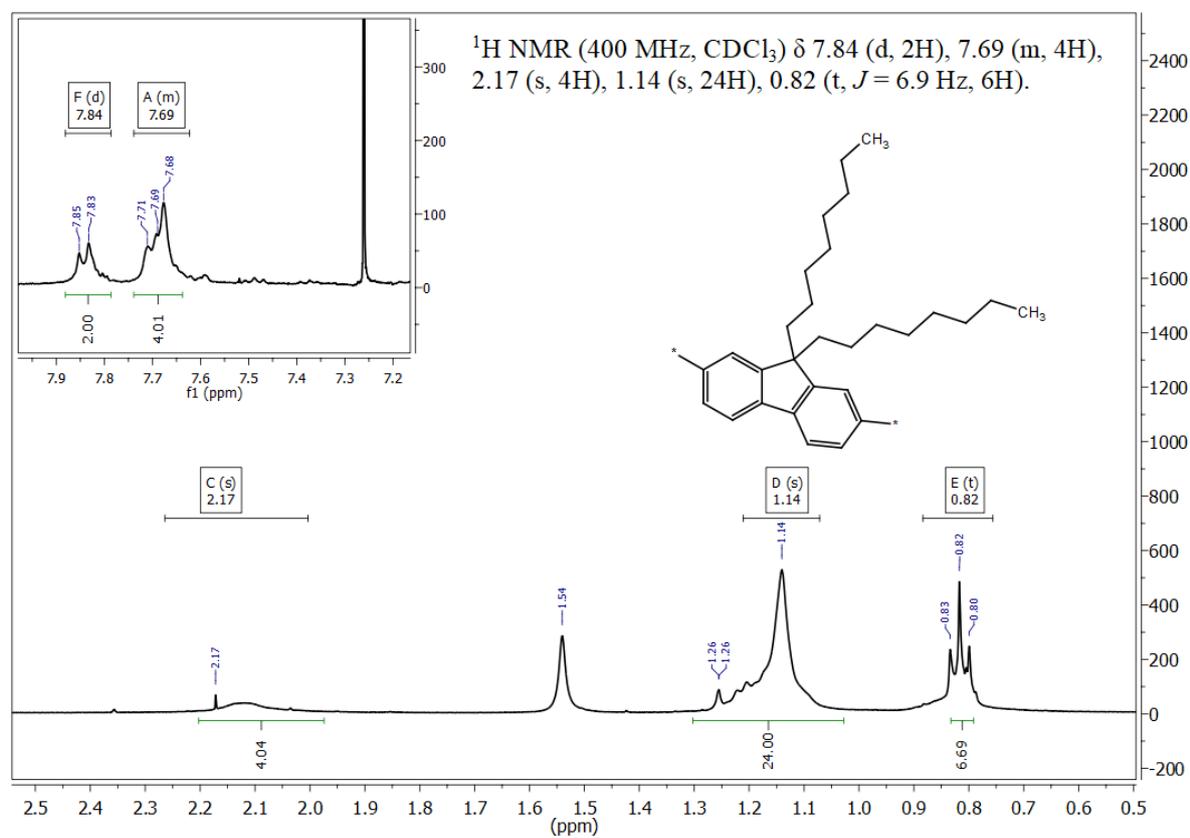

¹H NMR (400 MHz, CDCl₃) δ 7.84 (d, 2H), 7.69 (m, 4H), 2.17 (s, 4H), 1.14 (s, 24H), 0.82 (t, $J$ = 6.9 Hz, 6H).

**Figure S2** ¹H NMR spectrum of PFO synthesized in-house.

### 3.2. Size Exclusion Chromatography (SEC)

The synthesized polymers were analyzed by Size Exclusion Chromatography (SEC) to deduce their molecular characteristics, including molecular weights and dispersity (Đ) indices (Table S3). This analysis was performed using an 1100 Agilent 1260 Infinity device from Agilent Technologies. The SEC setup comprised an isocratic pump, autosampler, degasser, a thermostatic box for columns, and a differential refractometer MDS RI Detector. Data acquisition and processing were carried out using the Addon Rev. B.01.02 data analysis software from Agilent Technologies. Calibration for SEC-calculated molecular weight was accomplished using linear polystyrene standards ranging from 580 to 300,000 g/mol. The separation of analytes was achieved using a pre-column guard (5 μm, 50 × 7.5 mm) and two columns (PLGel 5 μm MIXED-C 300 × 7.5 mm and PLGel 5 μm MIXED-D 300 × 7.5 mm). $CHCl_3$ (HPLC grade) served as the solvent with a flow rate of 0.8 mL/min. The measurements were conducted at room temperature.

**Table S3** Molecular characteristics of the synthesized polymer batches measured by GPC.

| Acronym | Polymer no. | $M_n$ [g/mol] | $M_w$ [g/mol] | Đ |
|---|---|---|---|---|
| PFO-BPy | P1 | 8 300 | 11 650 | 1.40 |
| PFO | P2 | 23 600 | 53 400 | 2.26 |

### 3.3. Absorption spectroscopy

Optical absorption spectra of freshly collected supernatants were measured within the wavelength range of 280–1100 nm using a Hitachi U-2910 spectrophotometer. A double-beam mode was used with a pure solvent cuvette placed in the reference channel. The measurement was performed using 5 mm quartz cuvettes.

### 3.4. Photoluminescence excitation-emission mapping

Excitation-emission photoluminescent maps (PL) were acquired using a ClaIR microplate reader (Photonetc, Canada). The data were registered in the ranges of 460–900 nm (excitation) and 900–1600 nm (emission). The results were then visualized using OriginPro 2022b software.

### 3.5. Raman spectroscopy

Raman spectroscopy of drop-cast SWCNT samples was performed with a Renishaw inVia confocal Raman microscope in backscattering configuration and a 20x objective (Olympus MPlanFLN 20x Microscope Objective) equipped with a 532 nm (Renishaw, RL532-08) excitation laser. Over 200 individual spectra were acquired and averaged for each sample.

### 3.6. X-ray photoelectron spectroscopy

X-ray photoelectron spectroscopy (XPS) analysis was conducted in an ultra-high vacuum experimental setup (base pressure 8.5·10$^{-9}$ Pa) with the use of a PREVAC EA15 hemispherical electron energy analyzer fitted with a 2D-MCP detector. For sample excitation, a monochromated X-ray source (dual-anode PREVAC XR-40B source, RMC50 monochromator; Al-Kα excitation line with energy 1486.60 eV) was utilized. Survey spectra for initial sample characterization were recorded with pass energy (PE) set to 200 eV (scanning step of 0.8 eV). For high-resolution energy regions (scanning step of 0.07 eV), PE was set to 100 eV. To minimize analyzer-induced distortions, all measurements were performed with a normal take-off angle and the curved (0.8 x 25 mm) analyzer exit slit. The binding energy scale of the analyzer was calibrated to the Au 4f$_{7/2}$ (84.0 eV) region of the gold-covered sample, which was placed at the same sample stage [1]. To avoid possible charging-related parasitic effects, the binding energy scale was additionally referenced to carbon C1s, whose energy position was determined using the procedure proposed by Greczynski and Hultman [2]. Spectra were processed using CASA XPS® software (version 2.3.25) with the application of built-in algorithms. The particular components were fitted with a combination of Gaussian (30%) and Lorentzian (70%) functions. Shirley's function was applied for the background subtraction.

## 4. Modeling

### 4.1. DFT calculations

The spin-polarized Density Functional Theory (DFT) [3,4] calculations for (6,5) SWCNT, PFO-BPy6,6' of different lengths (1, 3, 5, or an infinite number of monomers) were carried out using a generalized gradient approximation (GGA) employing hybrid exchange and correlation functional, HSE06 [5,6] and double-ζ plus polarization numerical basis (DZP) sets, as implemented in SIESTA [7,8]. These calculations relied on our implementation of hybrid functionals within the SIESTA numerical package, where the Hartree–Fock exchange energy is evaluated by expanding numerical atomic orbitals (NAOs) as linear combinations of Gaussian-type orbitals (GTOs). Gaussian fitting was performed using the Levenberg–Marquardt algorithm. In our simulations, we used six Gaussian functions for both the first and second ζ of the s orbitals (for H, C, and N), five Gaussians for the first and second ζ of the p orbitals (for C and N), and four Gaussians for the first ζ of the d orbitals (for C and N). For hydrogen, which includes only a single-ζ p orbital in the DZP basis, five Gaussians were used. Calculations incorporating solvent effects via the COSMO continuum solvation model [9,10] were performed using the QuantumATK numerical package [11,12] using a medium basis set. The DFT calculations for finite polymers were performed in the 'molecule configuration' mode without any periodic boundary conditions being applied. The Brillouin zone was sampled only at the Γ point, while the density mesh cut-off for the real-space integrals was set at 350 Ry. For infinite polymers and SWCNTs, calculations were carried out in the 'bulk configuration' mode with 1D periodic boundary conditions applied. Consequently, the sampling of the Brillouin zone was increased to (1 × 1 × 3) and (1 × 1 × 7) k-points in the Monkhorst-Pack scheme [23]. All the structures were relaxed until the maximum force acting on any atom was lower than 0.004 eV/Å and the maximum stress changed by less than 0.001 GPa, while the self-consistent field (SCF) cycle was iterated until the total energy changed by less than $10^{-5}$ eV, and the density matrix elements differed by less than $10^{-5}$ per iteration. The solvent was not included in these calculations.

### 4.2. MD and MC calculations

To explore the interactions between [PFO-BPy6,6]$_5$ and (6,5) SWCNT, we conducted a series of molecular dynamics (MD) and time-stamped force-bias Monte Carlo (TFMC) simulations [13,14]. These simulations modeled infinite SWCNTs interacting with the polymer in toluene, acetonitrile, and DMSO solutions (Figure S7). Due to the use of three-dimensional periodic boundary conditions, the simulation boxes included only two units of the (6,5) SWCNT. The PFO-BPy6,6'$_5$ polymer was initially positioned near the SWCNT, aligned along its axis of symmetry. The simulation boxes, with dimensions of 4.7 nm × 4.7 nm × 8.4 nm along the X, Y, and Z axes, respectively, were large enough to prevent direct interactions between periodic images of the SWCNT–polymer complex. Each box was filled with one of three solvents using Packmol [15]: toluene (900 molecules), acetonitrile (2046 molecules), and DMSO (1196 molecules).

MD/MC calculations were performed using a full periodic table bonded valence forcefield - a Universal Force Field (UFF) potential [16] as implemented in QuantumATK [12,17]. Energy contributions to the UFF potential were represented by simple functions based on bond lengths, bond angles, torsion angles, inversion angles, and inter-atomic distances. The electrostatic interactions were calculated using the smooth-particle-mesh-Ewald (SPME) solver [18]. The cut-off used for calculating the real-space interactions was set to 7.5 Å, while the relative accuracy of SPME summation was set to 0.0001. Atomic partial charges on each atom were assigned using the QEq charge equilibration method [19]. Dispersive interactions were included in the form of a Lennard-Jones potential [20–22] with a 10 Å cut-off and a 2 Å smoothing length.

The production NPT simulations were preceded by short equilibration runs using the NPT ensemble and a Berendsen thermostat [23] at 300 K. Initial atomic velocities were randomly assigned based on the Maxwell-Boltzmann distribution. The thermostat relaxation time was set to 100 fs, and the simulations were performed with a time step of 0.02 fs over a duration of 5 ps (500,000 steps). These equilibration runs were followed by 20 ps (2,000,000 steps) of NPT simulations at 300 K and 1 bar, using a Martyna-Tobias-Klein barostat and thermostat [24]. The thermostat and barostat relaxation times were set to 500 fs and 1000 fs, respectively, with the time step maintained at 0.02 fs.

Following the production MD simulations, a brief geometry optimization was performed using the LBFGS algorithm for 10,000 steps [25]. The system was then equilibrated under microcanonical (NVE) conditions at 300 K for 5 ps (50,000 steps) using a 0.1 fs time step. This was followed by another round of geometry optimization: 2,000 steps using the first-FIRE algorithm [26] and a final 500-step L-BFGS minimization.

To extend the sampling and capture longer-timescale phenomena, we employed the time-stamped force-bias Monte Carlo (TFMC) method implemented in QuantumATK. TFMC simulations were conducted under NPT conditions using the Berendsen thermostat and barostat at 300 K and 1 bar, over a period of 5.31 ns (4,000,000 steps). The maximum atomic displacement per Monte Carlo step was set to 0.05 Å. The system compressibility was defined as 0.0005 bar$^{-1}$ and the barostat coupling factor was set to 1000.

The radial distribution functions (RDFs) and mass density profiles (MDPs) were calculated using the data obtained during the final 4 ns.

**4.3. DFTB calculations**

The density functional-based tight-binding (DFTB) method [27] was used to further analyze the differences in the interaction between PFO-BPy6,6' and the (6,5) SWCNT. Due to computational constraints, the systems used for DFTB modeling were reduced in size compared to those used in the MD/MC simulations shown in Figure S7. Specifically, smaller subsystems were extracted from the final MD/MC snapshots by retaining an ~8 Å thick solvent layer surrounding both the SWCNT and the polymer. These reduced configurations are illustrated in the insets of Figure 5. We also performed DFTB calculations for a reference system in which no polymer molecules were present (2 units of (6,5) SWCNT immersed in a toluene, acetonitrile, or DMSO solution). We used the Slater–Koster parameter set mio-1-1 [27–29] for C, N, and H atoms in systems containing toluene and acetonitrile, and the auorg-1-1 [27,28,30] parameter set for C, N, O, S, and H atoms in systems containing DMSO. The use of auorg-1-1 was necessary for DMSO, as the mio-1-1 parametrization resulted in incorrect molecular geometries. The DFTB calculations included a self-consistent charge correction that considered the charge fluctuations due to interatomic electron-electron interactions. Due to the size of the system, the Brillouin zone was sampled only at the Γ point. The density mesh cut-off was set to 10 Ha for real-space integrals, and the interaction maximum range was set to 10 Å. Systems were relaxed until the maximum force acting on any atom was lower than 0.05 eV/Å and the maximum stress changed by less than 0.1 GPa. The SCF cycle was iterated until the density matrix elements changed by less than $10^{-5}$ per iteration. During the DFTB calculations, 3D periodic boundary conditions were applied.

# 5. Results and discussion

## 5.1. Analysis of polymer-free SWCNTs modified with PhH in selected solvents

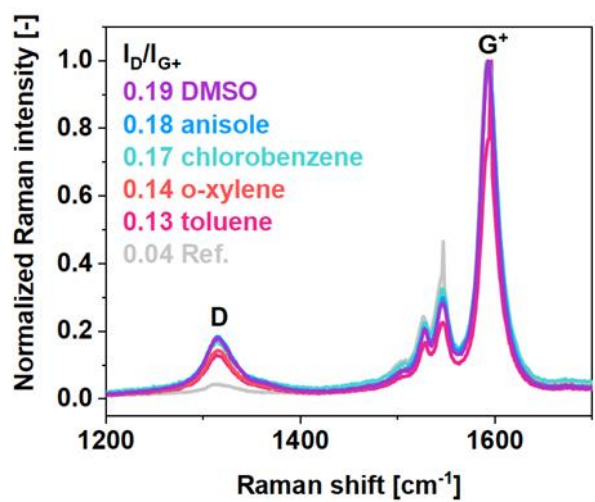

**Figure S3** Raman spectra of starting SWCNTs and SWCNTs covalently modified with PhH in toluene, o-xylene, chlorobenzene, anisole, and DMSO registered at the excitation wavelength of 532 nm.

## 5.2. Impact of basis sets on the spatial distribution of molecular orbitals in PFO-BPy

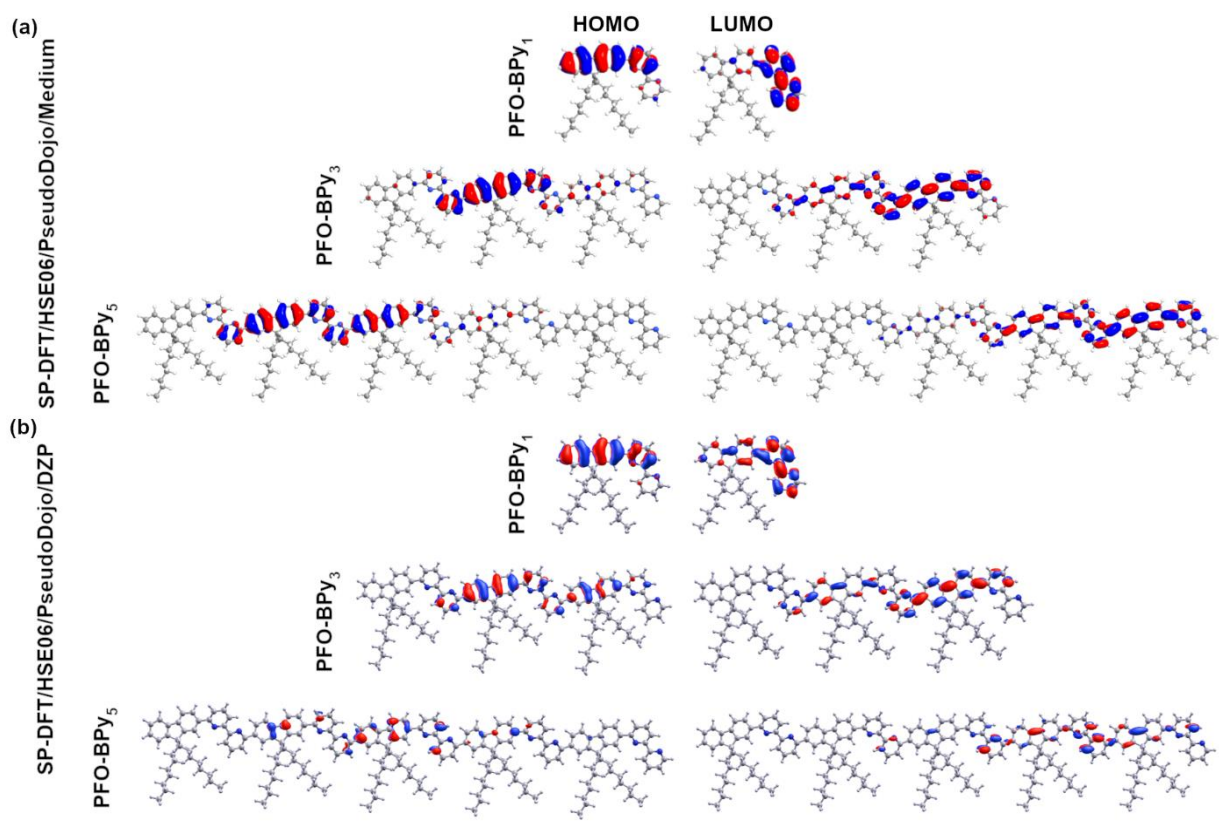

**Figure S4** HOMO and LUMO of fully relaxed PFO-BPy6,6 polymers of different lengths (1, 3, or 5 number of monomers) calculated at the (a) SP-DFT/HSE06/Medium level of theory using Quantum ATK visualized at a 0.06 Å$^{-3/2}$ isovalue and (b) SP-DFT/HSE06/DZP level of theory using SIESTA visualized at a 0.036 isovalue. Positive values of wave functions are represented by blue, whereas negative values are represented by red. Carbon, nitrogen, and hydrogen atoms are marked in gray, blue, and white, respectively. In agreement with previous studies [14,44], our DFT calculations showed that the PFO-BPy HOMO is mainly formed by the PFO unit and, to a lesser extent, by the BPy unit, whereas the PFO-BPy LUMO is localized only on the BPy unit.

## 5.3. Impact of solvent on the morphology and electronic properties of PFO-BPy

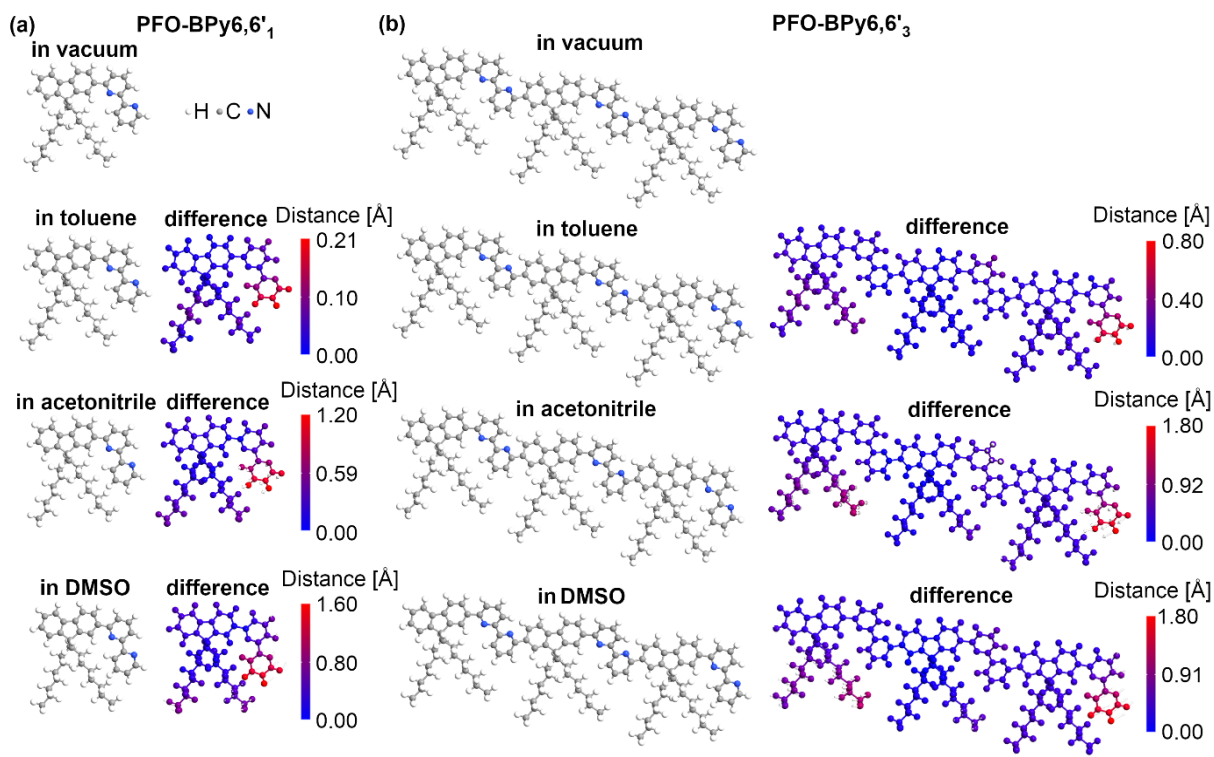

**Figure S5** Structural deviations between vacuum-optimized and solvent-relaxed geometries of PFO-BPy6,6' monomer (top) and trimer (bottom) calculated using the SP-DFT/HSE06/medium level of theory with the COSMO solvation model. Atom-wise displacements (in Å) relative to the vacuum configuration are shown for three solvents: toluene, acetonitrile, and DMSO. The most pronounced geometry changes are observed in the polar solvents.

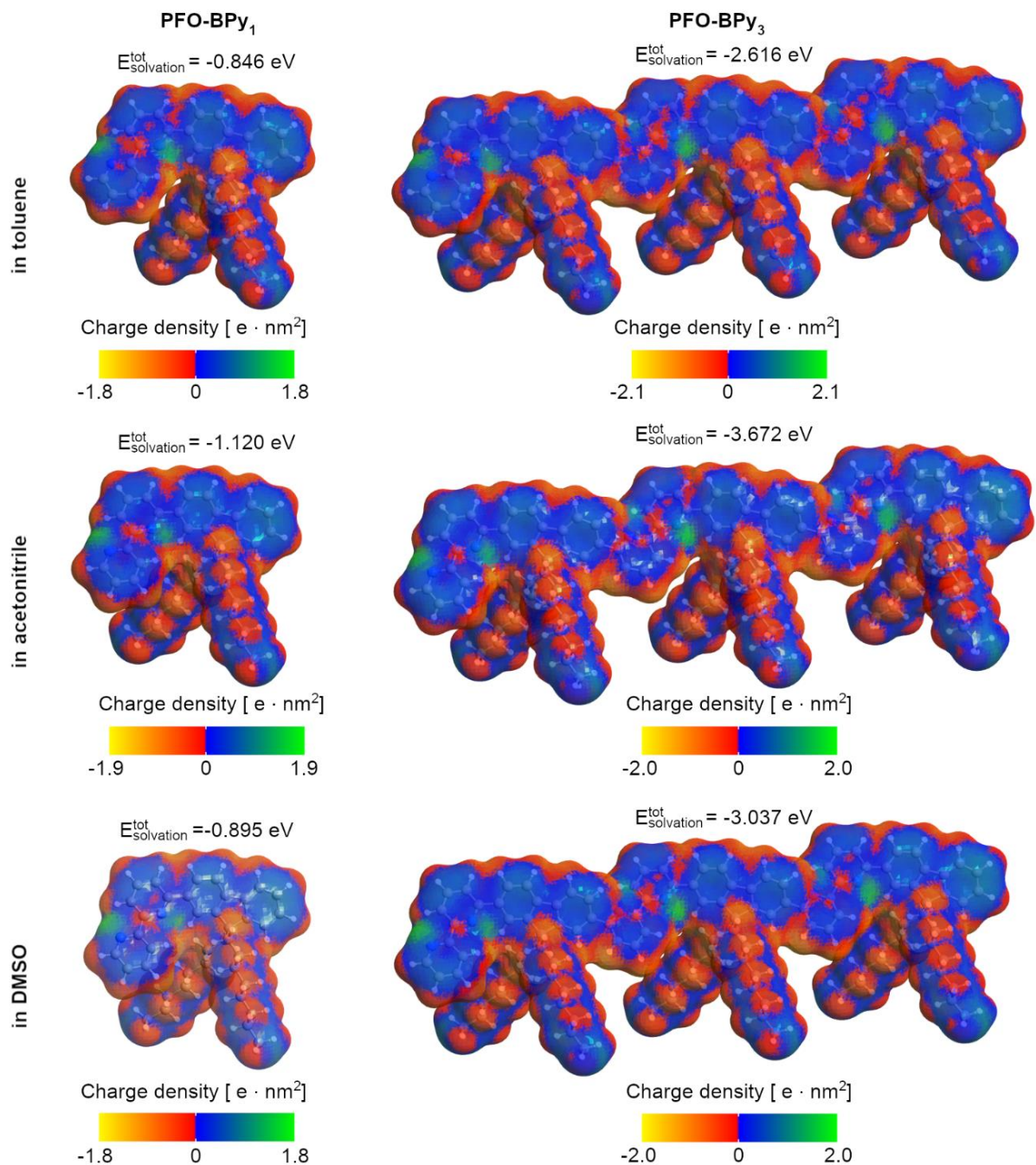

**Figure S6** Solvent-accessible surface charge distributions and corresponding solvation energies of PFO-BPy6,6' monomer (left) and trimer (right) calculated in toluene, acetonitrile, and DMSO using the SP-DFT/HSE06/medium level of theory with the COSMO solvation model.

## 5.4. Impact of solvent on the interaction between (6,5) SWCNT and PFO-BPy

The MD/MC modeling results presented in Figure S7 reveal subtle but consistent differences in the interactions between PFO-BPy6,6'$_5$ polymers and (6,5) SWCNTs in different solvent environments. Although the overall differences are relatively small, the radial distribution function (RDF) analysis indicates a slightly higher probability of the polymer being found in close proximity to the SWCNT surface when the system is immersed in toluene, compared to acetonitrile or DMSO (cf. the height of the first RDF peaks in Figure S7b, top panel).

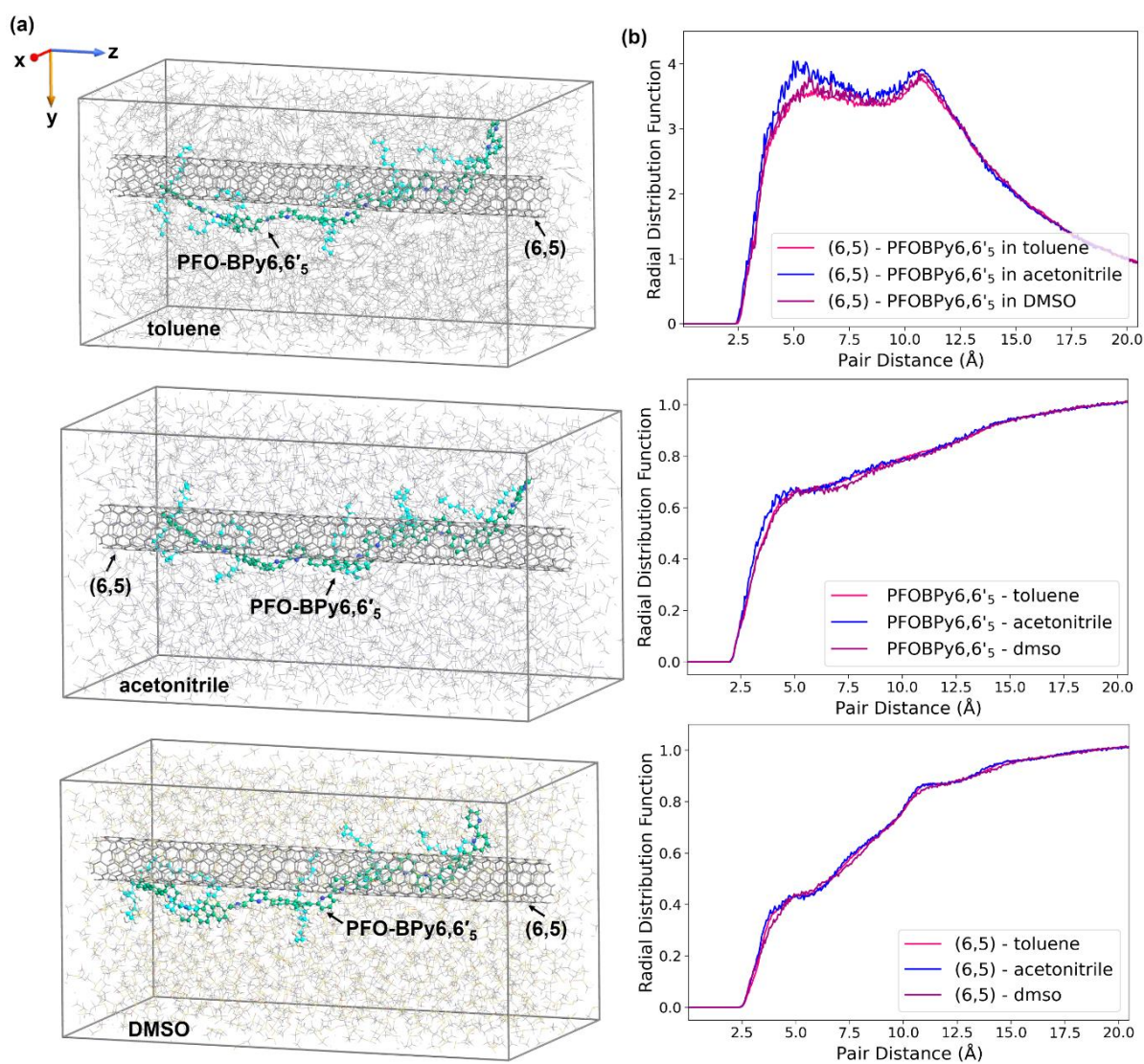

**Figure S7** MD/MC simulation of (6,5) SWCNT – PFO-BPy6,6'$_5$ interactions in solutions. (a) Snapshots of the final configuration for the simulation box containing two units of (6,5) SWCNT interacting with PFO-BPy6,6' (five monomers) in toluene (top panel), acetonitrile (middle panel), or in DMSO (bottom panel) solutions. SWCNTs, solvent, and polymer molecules are represented using stick, line, and ball-and-stick models, respectively. For clarity,

solvent molecules are drawn with a higher degree of transparency than other system components. Carbon atoms constituting the polymer backbone and side chain are marked in teal and cyan, respectively. Other carbon atoms are grey, whereas nitrogen, oxygen, sulfur, and hydrogen atoms are blue, red, yellow, and white, respectively. (b) RDFs of SWCNT-polymer (top panel), polymer-solvent (middle panel), and SWCNT-solvent atoms.

A more detailed RDF decomposition (Figure S8) shows that, in all solvents, the polymer backbone resides closer to the SWCNT surface than the side chains, suggesting preferential backbone–SWCNT interactions. Additional RDF analysis further indicates that toluene molecules tend to be slightly closer to both the polymer (Figure S7, middle panel) and the SWCNT surface (Figure S7, bottom panel) than molecules of acetonitrile or DMSO. A more accurate description of such interactions may require reactive force fields like ReaxFF or machine learning-based interatomic potentials, or higher-level methods such as DFTB or DFT, the results of which we present in the main text.

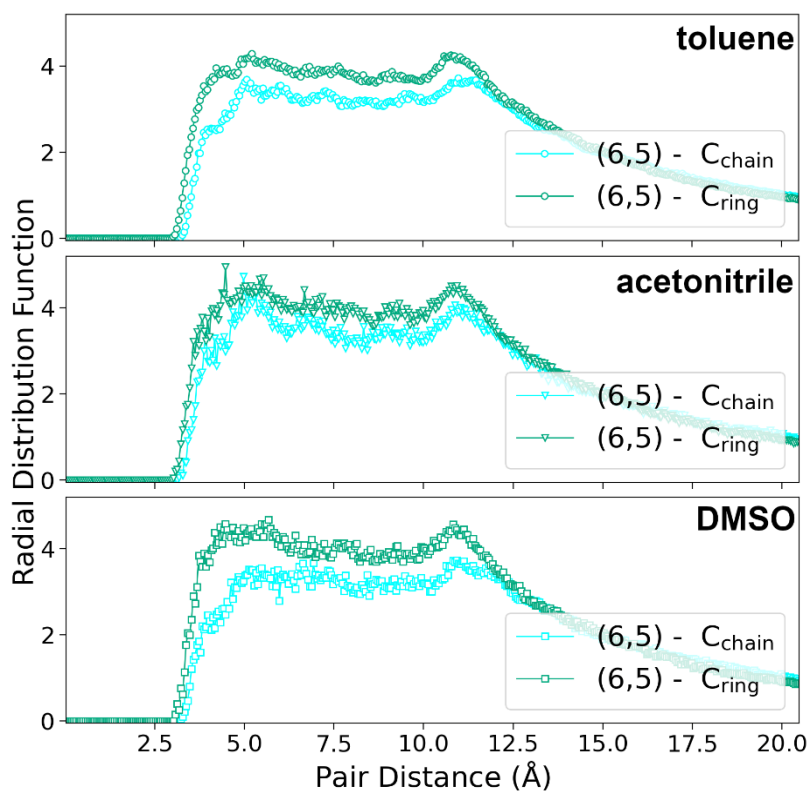

**Figure S8** RDFs of (6,5) SWCNT carbon-polymer backbone carbon ($C_{ring}$), SWCNT carbon-polymer side chain ($C_{chain}$) in three different solvents.

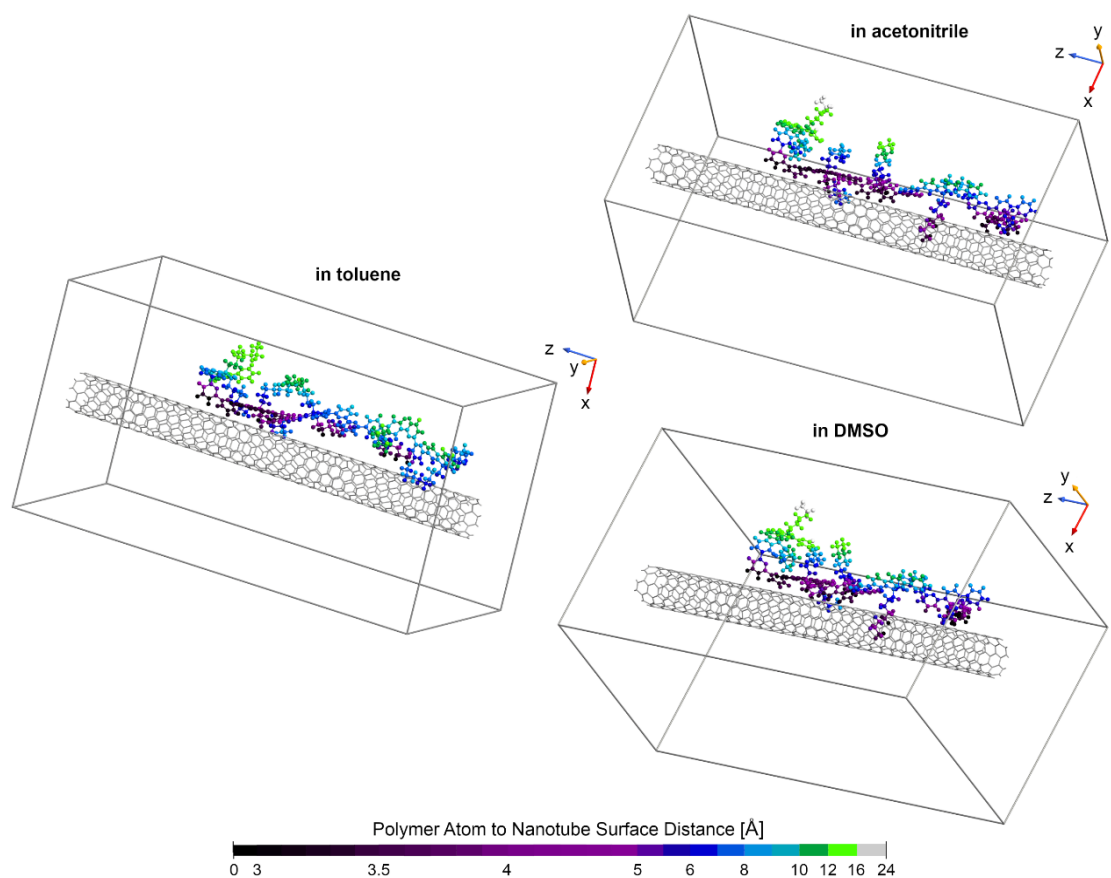

**Figure S9** Atomistic views of (6,5) SWCNT interacting with PFO-BPy6,6'$_5$ polymer in toluene, acetonitrile, or DMSO, obtained at the DFTB level. SWCNTs and polymers are shown using stick and ball-and-stick models, respectively. Solvent molecules are omitted for clarity. Polymer atoms are colored based on their distance from the SWCNT lateral surface, with darker shades indicating closer proximity and lighter shades indicating greater distance.